\documentclass[useAMS,usenatbib]{mn2e}
\usepackage{graphicx}
\usepackage{amsmath}
\usepackage{txfonts}
\usepackage{epsf, epsfig}
\usepackage{color}

\title[Electron re-acceleration in X-ray binary jets]{Electron transport with re-acceleration and radiation in the jets of X-ray binaries}
\author[Zhang et al. ]{Jian-Fu Zhang$^1$, Zhi-Ren Li$^1$, Fu-Yuan Xiang$^1$, Ju-Fu Lu$^2$\\
$^1$Department of Physics, Xiangtan University, Xiangtan, Hunan 411105, China; jfzhang@xtu.edu.cn\\
$^2$Department of Astronomy, Xiamen University, Xiamen, Fujian 361005, China}

\date{Accepted 2017 October 01. Received 2017 September 30; in original form 2017 June 12}

\begin{document}
\pagerange{\pageref{firstpage}--\pageref{lastpage}} \pubyear{2017}

\maketitle

\begin{abstract}
This paper studies acceleration processes of background thermal electrons in X-ray binary jets via turbulent stochastic interactions and shock collisions. By considering turbulent magnetized jets mixed with fluctuation magnetic fields and ordered, large-scale one, and numerically solving the transport equation along the jet axis, we explore the influence of such as magnetic turbulence, electron injections, location of an acceleration region, and various cooling rates on acceleration efficiency. The results show that (1) the existence of the dominant turbulent magnetic fields in the jets is necessary to accelerate background thermal electrons to relativistic energies. (2) Acceleration rates of electrons depend on magnetohydrodynamic turbulence types, from which the turbulence type with a hard slope can accelerate electrons more effectively. (3) An effective acceleration region should be located at the distance $>10^3R_{\rm g}$ away from the central black hole ($R_{\rm g}$ being a gravitational radius). As a result of acceleration rates competing with various cooling rates, background thermal electrons obtain not only an increase in their energies but also their spectra are broadened beyond the given initial distribution to form a thermal-like distribution. (4) The acceleration mechanisms explored in this work can reasonably provide the electron maximum energy required for interpreting high-energy $\gamma$-ray observations from microquasars, but it needs to adopt some extreme parameters in order to predict a possible very high-energy $\gamma$-ray signal.
\end{abstract}

\begin{keywords}
Black hole physics - radiation mechanisms: non-thermal - X-rays: binaries - acceleration of particles - turbulence
\end{keywords}

\section{INTRODUCTION}
A black hole X-ray binary is composed of a central black hole, a stellar companion, and an accretion disc surrounding the black hole, in which there is a source class producing bipolar relativistic radio jets that is called a microquasar (\citealt{Mirabel92}). It is generally believed that the bipolar relativistic extended radio jets are ejected from the central region around the compact object during the low/hard spectral state. For more than a decade, the multi-frequency radiative processes of microquasars have been widely investigating in the framework of  various theoretical models. There is a popular belief that synchrotron radiative processes of relativistic electrons injected in the jet can reproduce the radiation ranging from radio to infrared wavebands. However, the origin of both X-rays and soft $\gamma$-rays (e.g. MeV tail) is still actively debated. One possibility is that they are produced within the hot accretion flow by the Comptonization of soft X-rays from the accretion disc or by non-thermal processes in the hot corona that surrounds the accretion disc (\citealt{Esin01,Yuan05,Romero10a,ZAA12,Zhang13,Romero14}). Another possibility is that they originate from synchrotron radiation of relativistic electrons in the jet (e.g. \citealt{Markoff01,Markoff03}) or inverse Compton scattering (ICS) of the synchrotron and/or stellar (external Comptonization) photons (e.g. \citealt{Geor02,Kaufman02,Markoff05,Peer09,Peer12}).

From an observational point of view, four identified microquasars have been detected at the high-energy (GeV) $\gamma$-ray bands by the \emph{Fermi} Large Area Telescope (\emph{Fermi} LAT) and/or the AGILE satellite: Cygnus X-1 \citep[e.g.][]{Malyshev13,Zanin16}, Cygnus X-3 \citep[e.g.][]{Abdo09a,Tavani09}, SS433 \cite[][]{Bordas15}, and V404 Cygni (\citealt{Loh16,Piano17}). The former three sources are classified as high-mass microquasars according to their companion star mass. The latter is a low-mass microquasar, from which the high-energy $\gamma$-ray emission was detected by AGILE in the 50--400~MeV energy ranges with a detection significance of $4.3\sigma$. This observational result is in agreement with a contemporaneous observation by \emph{Fermi} LAT, which therefore strengthens the reliability of the observation obtained in \cite{Piano17}. However, no significant $\gamma$-ray emission signal has been reported above 400 MeV so far.

At the very high-energy (TeV) bands, three sources  of microquasar candidates have been detected by the Major Atmospheric Gamma Imaging Cherenkov Telescope (MAGIC), the High Energy Stereoscopic System (HESS) and/or the Very Energetic Radiation Imaging Telescope Array System (VERITAS):  LS 5039 \citep[][]{Aharonian05}, LS I +$61^{\circ}$ 303 \citep[][]{Albert06}, and Cygnus X-1 \citep[][]{Albert07}. The classification of the former two sources is subjected to a vigorous debate; one now is inclined to consider them as pulsar binary systems powered by the rotational energy of the central pulsars (e.g. \citealt{Dubus06,Dubus13}, but see \citealt{Romero07}). At present, these two sources are detected also in GeV $\gamma$-ray bands (\citealt{Abdo09b,Abdo09c}). The latter is a confirmed high-mass microquasar source. Since its flare radiation was observed for a few hours with a significance of $4.0\sigma$, many follow-up studies have been carried out but a positive radiation signal has not been re-detected . Therefore, the very high-energy observations of Cygnus X-1 remains to be confirmed.

The study of the origin of high-energy and very high-energy $\gamma$-ray radiation is a very active topic. In the framework of hadronic and/or leptonic models, many theoretical studies are committed to a jet dissipation region, in which relativistic particles injected may produce broadband photon radiation (e.g. \citealt{Romero03,Bosch06,Khangulyan08,Romero08,Zhang11,Vila12,ZAA14,Zhang14}) ranging from radio to TeV $\gamma$-ray bands, or high-energy neutrino radiation (e.g. \citealt{Reynoso09,Zhang10}), but so far there has been no consistent conclusion. We notice that almost all of the works in the literature consider relativistic particles (electrons and protons) as a direct injection source, that is, one assumes in prior that the relativistic particles have been accelerated by some unknown mechanisms. In other words, the previous works did not determine a detailed acceleration mechanism of relativistic particles, which leads to no self-consistent results between the particle distributions and emission output spectra.

On the basis of the description given above, we know that a positive signal at the TeV bands remains to be detected in microquasars. The upper limits of photon energies detected only from low-mass microquasar V404 Cygni are 400~MeV. Therefore, we want to ask that whether a microquasar system can indeed produce observable TeV $\gamma$-ray radiation and how the acceleration efficiency of relativistic particles is in the jet of microquasars. The purpose of the paper is to study re-acceleration processes of relativistic particles, and to explore the influence of magnetic turbulence, electron injections and various cooling channels on the acceleration of the electrons. In this regard, our studies can alleviate no self-consistent of the previous works between output photon spectra and injected particle spectra.

This paper is organized as follows. We present the basic properties of the particle acceleration model in the next section and the description of numerical methods in Sect. 3. In Sect. 4, we focus on studying the influence of such as accelerator location, the properties of magnetic turbulence, and various cooling rates on the acceleration processes. The discussions and the summary of the paper are provided in Sects. 5 and 6, respectively.

\section{MODEL DESCRIPTION}
\begin{figure}
\centerline{
\includegraphics[width=15.0cm]{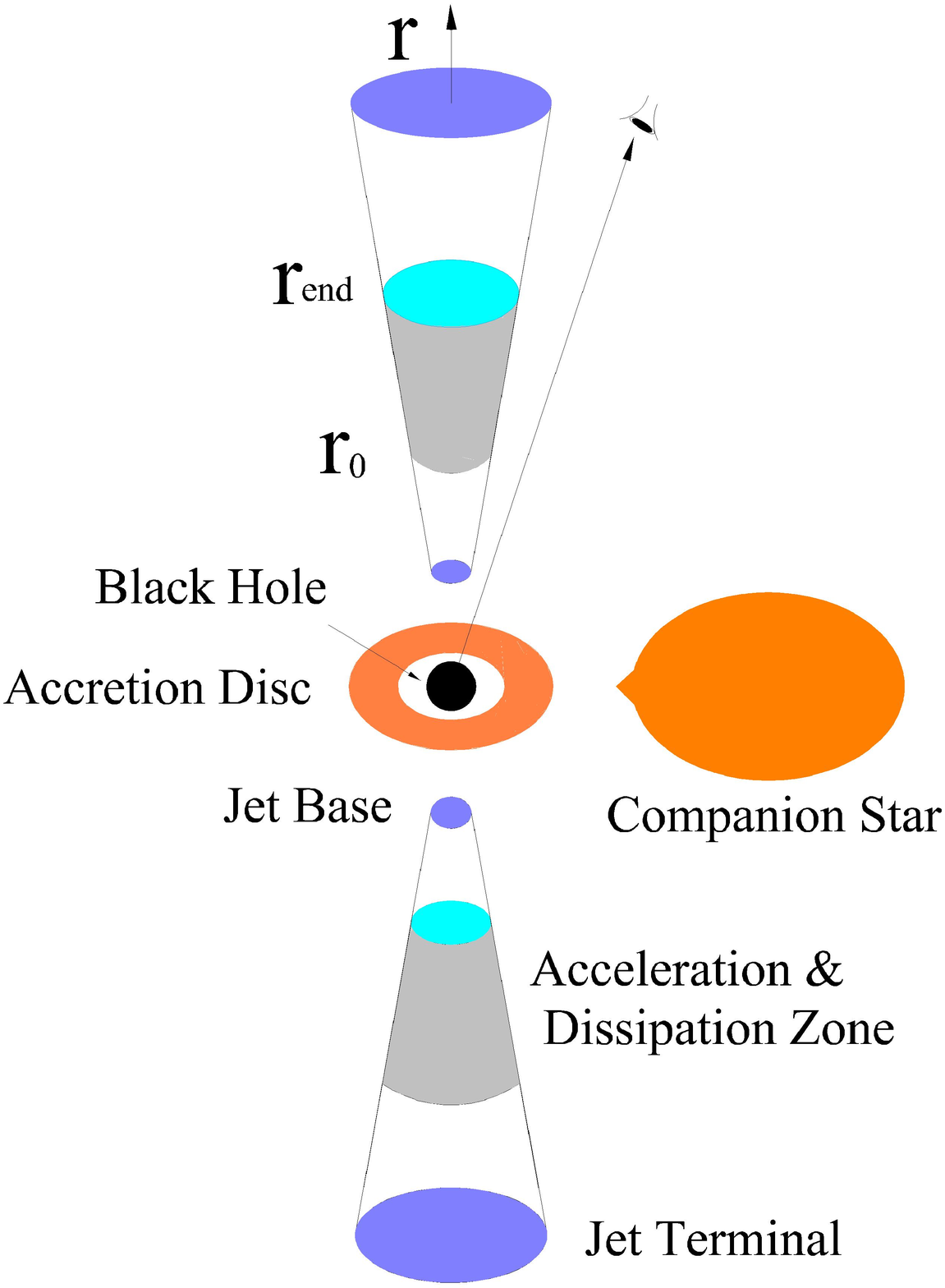}
}
\caption[ ]{Sketch of a black hole X-ray binary.  An acceleration zone with radiative dissipation in the jet is constrained to the region between $r_{\rm 0}$ and $r_{\rm end}$ . } \label{fig:Sketch}
\end{figure}

This paper adopts a conical structure of the turbulent magnetized X-ray binary jet to explore how relativistic particles are accelerated in the jet (see Fig. \ref{fig:Sketch}). Specifically, we restrict ourselves to a leptonic scenario, which has an isotropic distribution of pitch-angles, with the goal of investigating leptonic transport, acceleration and radiation in the jets\footnote{It is noticed that particle transport in a magnetized plasma corona of X-ray binaries is studied in \cite{Vieyro12}, and in the other scenarios such as supernova blast waves, accretion shocks and solar flares (\citealt{Jokipii87}). }.  We emphasize that the methods we describe in this work can be applied to more heavy particles, which can also produce broadband radiation by proton-proton and/or proton-$\gamma$ interactions (e.g. \citealt{Romero03,Romero08}).

The kinetic equation governing the evolution in the jet of an ensemble of electrons is given by (e.g. \citealt{Schlickeiser02})
\begin{equation}
\triangledown'\cdot (\mathbf{\upsilon} f-\kappa\triangledown' f)={1 \over p^2}{\partial \over \partial p}\left[p^2\left(D {\partial f \over \partial p} + {\triangledown' \cdot \mathbf{\upsilon} \over 3}pf + {dp\over dt'}f\right)\right]+Q(p,r') \label{EQorigin}
\end{equation}
in the jet frame of the reference. The terms on the left-hand side correspond to both spatial convection and diffusion along the jet, which indicate the transport process of electrons. The first and second terms on the right-hand side are momentum diffusion and convection, respectively, which are associated with the acceleration process of electrons. In equation (\ref{EQorigin}), $dp/dt' = (dp/dt')_{\rm rad} + (dp/dt')_{\rm gain}$ describes both a radiation energy loss rate and a gain rate (due to shock collision) of electrons\footnote{In this paper, the jet co-moving and observer references correspond to a zero jet bulk velocity and $V_{\rm jet}$, respectively. The quantities in the jet reference are denoted by primed symbols, when it is necessary to distinguish between them in both frames. }, with the proper time $dt'=dt/\Gamma_{\rm j}$; the latter corresponds to the first-order \emph{Fermi} acceleration. $\kappa$ and $D$ are spatial and energy diffusion coefficients, respectively. Furthermore, $Q$ is the energy injection spectrum of low-energy background thermal electrons and $\Gamma_{\rm j}$ is the bulk Lorentz factor of the jet. It should be noticed that the time term has been omitted in equation (\ref{EQorigin}) and our works are dealing with a steady state case. In other words, this equation cannot be used to study flares or variability (see \citealt{Romero17} for more discussions).

For the sake of simplicity, one usually replaces the spatial diffusion term (e.g. the second term on the left-hand of equation \ref{EQorigin}) with a simplified escape term (e.g. \citealt{Petrosian12}), which is assumed as an energy independent diffusion of electrons from the system. In
the turbulent physical environment of X-ray binary jets, the contribution of spatial diffusion term of electrons is negligible (e.g. \citealt{Khangulyan08}). Therefore, in the case of an isotropic distribution form of electrons' momenta, the electron distribution in the momentum space
is written as $\tilde{N}dp=4\pi p^2fdp$, and then equation (\ref{EQorigin}) can be written as
\begin{equation}
{1\over r^2}{\partial \over \partial r}[\Gamma_{\rm j}\beta_{\rm j}cr^2\tilde{N}]={\partial\over\partial p}\left[D{\partial \tilde{N}\over\partial p} -\left({2D\over p} + {\triangledown \cdot \upsilon \over 3}p +{\Gamma_{\rm j}dp \over dt}\right)\tilde{N} \right]+\tilde{Q}(p,r)\label{EQtilde}
\end{equation}
in the observer frame. The above equation is derived based on a spherical coordinate (with components $r$, $\theta$ and $\phi$); we assumed that the distribution function of electron $\tilde{N}$ depends spatially only on the radial coordinate $r$.  The magnetic turbulence interaction provides both stochastic (first $D$ term) energy diffusion (gain) and systematic (second $D$ term) energy gain. The term $\triangledown \cdot \upsilon/3$ is referred to as an adiabatic loss rate; hereafter, we include this term into the term $dp/dt$ and rewrite as $dp/dt = (dp/dt)_{\rm rad} + (dp/dt)_{\rm gain} + (dp/dt)_{\rm adi}$ in the observer frame. The source term $\tilde{Q}(p,z)$, i.e. the injection rate of background electrons, has the dimension $\rm cm^{-3}\ erg^{-1}\  s^{-1}$.

In order to make equation (\ref{EQtilde}) more compact and convenient for a numerical treatment, we let $N(p, r)=\tilde{N}\pi R_{\rm jet}^2\Gamma_{\rm j}\beta_{\rm j}ct$ in units of counts $\rm erg^{-1}$, and $Q(p, r)=\tilde{Q}\pi R_{\rm jet}^2\Gamma_{\rm j}\beta_{\rm j}ct$ in units of counts $\rm erg^{-1}\ s^{-1}$. Here, $\beta_{\rm j}$ is the bulk velocity, and $R_{\rm jet}=r{\rm tan} \varTheta$ is a radius of the jet at the height $r$, where $\varTheta$ is the half-opening angle of the jet. Hence, by using the relation $dr=\beta_{\rm j}cdt$, equation (\ref{EQtilde}) is rewritten as
\begin{equation}
{\partial N(p,r)\over\partial r}={\partial \over \partial p}\left[A{\partial \over \partial p}N(p,r)+BN(p,r)\right]+{Q(p,r)\over \Gamma_{\rm j}\beta_{\rm j}c},\label{dNdz}
\end{equation}
where, $A=D/\Gamma_{\rm j}\beta_{\rm j}c$, $B=-2A/p+dp/dr$. The energy gain plus loss rate of relativistic electrons along the jet is
\begin{equation}
{dp\over dr}=\left(dp\over dr\right)_{\rm adi}+\left(dp\over dr\right)_{\rm rad}+\left(dp\over dr\right)_{\rm gain}, \label{dpdr}
\end{equation}
with $(dp/dr)_{\rm rad}=(dp/dt')_{\rm rad}/\Gamma_{\rm j}\beta_{\rm j}c$ and $(dp/dr)_{\rm adi}=-2p/3r$ for a lateral (two-dimensional) expansion jet.
The total radiative loss rate of relativistic electrons, $(dp/dt')_{\rm rad}$, include synchrotron, ICS components in the Klein Nishina regime (\citealt{Moderski05}) from the synchrotron soft photon, companion star, accretion disc (see \citealt{Zhang14} for more details) and background thermal photons. For the latter, we assume that background photon energy density is equal to the local magnetic energy density,  and these photons present a Maxwell distribution with characteristic energy of $T_{\rm e}$. In addition, we also consider another radiative cooling from relativistic bremsstrahlung radiative process, which is given by (\citealt{Blum70})
\begin{equation}
\left(dp\over dt'\right)_{\rm brem}=-4r_{\rm e}^2\Xi cn'_{\rm H}\left[{\rm log}(2p)-{1\over3} \right]p.
\end{equation}
Here, $r_{\rm e}$ and $\Xi$ are the classical electron radius and the fine structure constant, respectively, and $n'_{\rm H}$ is the thermal proton number density (see equation \ref{numden}) .  The term $(dp/dr)_{\rm gain}$ in equation (\ref{dpdr}) corresponds to the first-order \emph{Fermi} acceleration rate due to a shock interaction, i.e.  $(dp/dr)_{\rm gain}=(dp/dr)_{\rm sh}$, which will be described below.

Magnetic fields mixed with both a turbulent ($\delta B$) and an ordered ($B$) components are included into the turbulent magnetized jet matter. The interactions of background electrons in the jets with magnetohydrodynamic (MHD) turbulence result in both a first-order \emph{Fermi} acceleration when a shock is produced by a compression of the ejected matter from the disc-corona region, and a stochastic (second-order \emph{Fermi}) acceleration when random magnetic turbulence exists. In this work, we focus on acceleration in the region of Alfv{\'e}nic turbulence, in which turbulent Alfv{\'e}n waves energize low-energy electrons via resonant pitch angle scattering. The spectral distribution of magnetic turbulence is expressed as $W(k)\propto k^{-\alpha}$ in terms of the wave number $k=|\textbf{k}|=2\pi/\lambda$, where, $\lambda$ is the wavelength of the plasma waves.  The spectral index $\alpha$ is between 1 and 2. $\alpha=1$ would imply infinite energy content, and $\alpha=2$ corresponds to the case of diffusive shock waves, $\alpha=5/3$ the Kolmogorov spectrum, $\alpha=3/2$ the Kraichnan spectrum (\citealt{Zhou90}).  One could thus obtain the turbulent energy density $\delta B^2/8\pi=\int W(k)dk$ in the inertia range of magnetic turbulence.

Turbulent MHD waves is an ideal stochastic scattering agent for an energetic process of relativistic particles. For an isotropic pitch-angle scattering, the diffusion coefficient is given by (\citealt{Dermer96})
\begin{equation}
D={\pi \over 2}\left[{\alpha-1 \over \alpha (\alpha+2)}\right]\zeta\beta_{\rm A}^2(ck_{\rm 0})(R_{\rm NL}k_{\rm 0})^{\alpha-2}{p^\alpha\over \beta},\label{Dpp}
\end{equation}
where $\beta_{\rm A}=V_{\rm A}/c$ and $V_{\rm A}$ is the velocity of Alfv{\'e}n waves, $\zeta=\delta B^2/B^2$ is the ratio of turbulent field energy density to the ordered one, $R_{\rm NL}=m_{\rm e}c^2/eB$ is the non-relativistic Larmor radius of the electrons, $\beta=p/\gamma$ is the dimensionless velocity of the electrons with $p=\sqrt{\gamma^2-1}$, $k_{\rm 0}=2\pi/\lambda_{\rm max}$ is the minimum wavenumber of MHD waves, and $\lambda_{\rm max}\simeq R_{\rm jet}$ is the maximum wavelength of the Alfv{\'e}n waves.  As seen in equation (\ref{Dpp}), the momentum diffusion coefficient reflects the power-law behavior of the magnetic turbulence spectrum $D\propto p^\alpha$. Obviously, the usual Bohm diffusion, $D\propto p$, is a limiting case of the currently used diffusion behavior. The systematic energy gain rate is associated with the diffusion coefficient $D$ by (e.g. \citealt{Dermer96})
\begin{equation}
{d\gamma \over dt'}={1\over p^2}{\partial\over p}[p^2\beta D]. \label{SEG}
\end{equation}
On the basis of equations (\ref{SEG}) and (\ref{Dpp}), we obtain the  energy gain rate along the jet as follows
\begin{equation}
\left({dp\over dr}\right)_{\rm tur}\approx{1\over\Gamma_{\rm j}\beta_{\rm j}c}{\pi\over 2}{\alpha-1 \over \alpha}\zeta\beta_{\rm A}^2(ck_{\rm 0})(R_{\rm NL}k_{\rm 0})^{\alpha-2}{p^\alpha\over \beta}.\label{dgdztur}
\end{equation}

As a result of interactions of electrons with a random fluctuation field of MHD waves, these electrons can repeatedly cross the shock front propagating along the jet, which leads to the another scenario, i.e. occurrence of the first \emph{Fermi} acceleration process. In the frame of the co-moving jet, the upstream and downstream flows of the shock are non-relativistic, which allows us to use a classical description of particle accelerations in the case of a non-relativistic shock (e.g. \citealt{Bell78}). The mean energy gain rate along the jet axis is (e.g. \citealt{Brunetti04})
\begin{equation}
\left({dp\over dr}\right)_{\rm sh}={1\over\Gamma_{\rm j}\beta_{\rm j}c}{u^2_{-}\over \chi}\left({\chi-1\over \chi+1}\right) {p\over 3\kappa}\label{dgdzsh}
\end{equation}
in the laboratory reference frame, where $\chi=u_-/u_+$ is the shock compression ratio ($\chi=4$ for a strong shock), with $u_-$ and $u_+$ denoting the upstream and downstream velocity in the frame of the shock, respectively. $\kappa=c\ell_{\rm MHD}/3$ is the spatial diffusion coefficient; here $\ell_{\rm MHD}$ is the coherence length of MHD waves approximately equal to the mean-free path $L$. According to \cite{OSullivan09}, the mean-free path is given by
\begin{equation}
L=\zeta R_{\rm L}^{2-\alpha}/\lambda^{1-\alpha}_{\rm max},
\end{equation}
where, $R_{\rm L}=m_{\rm e}c^2\gamma\beta/eB$ is the relativistic Larmor radius of the electrons.

As for the turbulent magnetized jet, we consider that its dynamic structures are dominated by the thermal proton component. Based on this assumption, the jet matter energy density is expressed as (\citealt{Bosch06})
\begin{equation}
U'_{\rm H}(r)={\dot{m}_{\rm jet}\over \pi R_{\rm jet}^2V_{\rm jet} m_{\rm H}}\times{1\over 2} m_{\rm H}V_{\rm exp}^2={\dot{m}_{\rm jet}\over 2\pi r^2}V_{\rm jet},
\end{equation}
where, $m_{\rm H}$ is the proton mass, and $V_{\rm exp}=V_{\rm jet}{\rm tan}\varTheta$ is lateral expansion velocity with the jet bulk speed $V_{\rm jet}=\beta_{\rm j}c$. The jet matter ejection rate, $\dot{m}_{\rm jet}=\eta_{\rm jet}\dot{m}_{\rm acc}$, accounts for a fraction of the accretion rate of the compact object, $\dot{m}_{\rm acc}$. The number density of the thermal protons in the jet is (e.g. \citealt{Reynoso09})
\begin{equation}
n'_{\rm H}(r)={\eta_{\rm jet}\dot{m}_{\rm acc}c^2\over \pi R_{\rm jet}^2V_{\rm jet} m_{\rm H}c^2}. \label{numden}
\end{equation}
The ordered large-scale magnetic field in the jet co-moving frame at different distances from the central compact object is defined by
\begin{equation}
B'(r)=\sqrt{\xi 8\pi U'_{\rm H}(r)},\label{magfield}
\end{equation}
where, $\xi$ is the ratio of the ordered magnetic energy density to the matter energy one. On the basis of equations (\ref{numden}) and (\ref{magfield}), one could obtain the Alfv{\'e}n velocity
\begin{equation}
V_{\rm A}={B'(r)\over \sqrt{4\pi n'_{\rm H}m_{\rm H}}}.\label{Alfel}
\end{equation}
In the following, we will explore different injection sources of electrons in Sect. \ref{NumRes}, which is assumed to be a function form of $Q(p,r)=Q_{\rm 0}(r)F(p)$. The normalization constant of the electrons, $Q_{\rm 0}(r)$, is determined by
\begin{equation}
Q_{\rm 0}(r)=L_{\rm e}\times[\int F(p)pdp]^{-1},
\end{equation}
at the distance $r$ from the central black hole, where, $L_{\rm e}=\eta_{\rm e}\eta_{\rm jet}\dot{m}_{\rm acc}c^2$ is the power of injection electrons, accounting for a fraction $\eta_{\rm e}$ of the jet power.

\section{NUMERICAL METHODS}

The finite difference numerical technique proposed by \cite{Chang70} is used to solve equation (\ref{dNdz}). The methods of
solving this equation is similar to those presented in \cite{Park96} and \cite{Donnert14} for a time-dependent
Fokker-Planck equation. The difference is that the spatial evolution $r$ (rather than a time evolution $t$) along the axis is involved in our work.

We adopt the logarithmic spatial and momentum grids with $N+1$ points. The discrete spatial steps are indicated by $\triangle r=r_{j+1}-r_j$ and
momentum steps by $\triangle p=p_{i+1}-p_i$ . The midpoint between two momentum mesh points is defined by an arithmetic mean $p_{i+1/2}=(p_{i+1}+p_i)/2$, and the midpoint difference is defined by $\triangle p_{i+1/2}=p_{i+1}-p_i$, thus
we have $\triangle p_{i}=(p_{i+1}-p_{i-1})/2$. Equation (\ref{dNdz}) is differentiated in terms of fluxes as
\begin{equation}
{N_i^{j+1}-N_i^j\over \triangle r}={F_{i+1/2}^{j+1}-F_{i-1/2}^{j+1}\over \triangle p_i} + {Q_i\over \Gamma_{\rm j}\beta_{\rm j}c },\label{EQDisct}
\end{equation}
for $i$ and $j\in [0,N]$. It should be noted that the subscript $\rm j$ of both $\Gamma_{\rm j}$ and $\beta_{\rm j}$ indicate an abbreviation of the term `\emph{jet}' rather than a parameter. The flux function
\begin{equation}
F=A{\partial \over\partial p}N(p,r)+BN(p,r),
\end{equation}
is discretized as (\citealt{Park96})
\begin{align}
F_{i+1/2}^{j+1}=(1-\varPhi_{i+1/2})B_{i+1/2}N_{i+1}^{j+1}+\varPhi_{i+1/2}B_{i+1/2}N_{i}^{j+1}+A_{i+1/2}{N_{i+1}^{j+1}-N_i^{j+1}\over\triangle p_{i+1/2}}\nonumber \\
={A_{i+1/2}\over \triangle p_{i+1/2}}[\varPsi^+_{i+1/2}N^{j+1}_{i+1}-\varPsi^-_{i+1/2}N_i^{j+1}],\label{FluxFunc}
\end{align}
and $F_{i-1/2}^{j+1}$ can be done in a similar way. $\varPhi_i$ and $\varPsi_i^{\pm}$ in equation (\ref{FluxFunc}) is defined by
\begin{equation}
\varPhi_i={1\over\varPi_i}-{1\over e^{\varPi_i}-1}, \ \ \ \ \varPi_i={B_{i+1/2}\over A_{i+1/2}}\triangle p_{i+1/2},
\end{equation}
\begin{equation}
\varPsi_i^+={1\over 1-e^{-\varPi_i}}, \ \ \ \ \varPsi_i^-={1\over e^{\varPi_i}-1},
\end{equation}
according to \cite{Park96}. Substituting equations (\ref{FluxFunc})
into (\ref{EQDisct}), we obtain a tridiagonal matrix of linear equations
\begin{equation}
-a_i N^{j+1}_{i-1}+b_iN_i^{j+1}-c_iN_{i+1}^{j+1}=e_i.
\end{equation}
where, the coefficients are given by
\begin{equation}
a_i={\triangle r\over \triangle p_i}{A_{i-1/2}\over \triangle\ p_{i-1/2}}\varPsi^-_{i-1/2},
\end{equation}
\begin{equation}
b_i={\triangle r\over \triangle p_i}{A_{i+1/2}\over \triangle p_{i+1/2}}\varPsi^+_{i+1/2},
\end{equation}
\begin{equation}
c_i=1+{\triangle r\over \triangle p_i} \left[{A_{i-1/2}\over \triangle p_{i-1/2}}\varPsi^+_{i-1/2}+ {A_{i+1/2}\over \triangle p_{i+1/2}}\varPsi^+_{i+1/2}\right],
\end{equation}
\begin{equation}
e_i=N^j_i+{\triangle r\over \Gamma_{\rm j}\beta_{\rm j}c}Q_{i}.
\end{equation}
Assuming the no-flux boundary conditions, that is,
\begin{equation}
a_0=c_N=0,
\end{equation}
which can ensure a positivity and a conservation of electron numbers, we can solve the tridiagonal matrix by using the numerical procedures in \cite{Press01}.

\section{Numerical Results}
\label{NumRes}
In this section, we use some common parameters of a general microquasar to explore the relativistic electron acceleration and radiative processes
via both turbulent stochastic and diffusive shock acceleration mechanisms. These parameters are given as follows: a black hole mass of $M_{\rm BH}=10M_{\rm \sun}$, a distance of $d=2\ \rm kpc$ to the Earth, an effective surface temperature of $T_{\rm s}=6\times10^4\ \rm K$ and a radius of $R_{\rm s}=10R_{\rm \sun}$ of the surrounding companion star, an orbital radius of $R_{\rm or}=10^2R_{\rm \sun}$ of binary system, an accretion rate of $\dot{M}_{\rm acc}=10^{-8}M_{\rm \sun}\ \rm yr^{-1}$ of the central black hole, viewing angle of $\varphi=30^\circ$ of an observer, a bulk Lorentz factor of $\Gamma_{\rm j}=2$ of the jet, and a half-opening angle with $\varTheta=5^\circ$ of the jet (e.g. \citealt{Romero08,Zhang09}). With the above relevant parameters as well as setting $\eta_{\rm jet}=0.1$ and $\xi=1$, we obtain the magnetic field strength to be $\sim10^6\ \rm G$ at the base of the jet, which is approximately equal to  $50 R_{\rm g}$; here, $R_{\rm g}=GM_{\rm BH}/c^2$ is the gravitational radius of the black hole. Furthermore, the maximum amplitude of the magnetic fluctuation is to be $\sim 10^3 \rm G$ within the acceleration zone, provided that we assume the ratio of the turbulent magnetic energy density to the ordered one as $\zeta=1$ and the inner boundary of acceleration zone as $10^4R_{\rm g}$.

Besides, the characteristic temperature of background thermal electrons is considered as a low value of $0.001$ in units of $m_{\rm e}c^2$ for the sake of simplicity (see also \citealt{Aharonian17} for the physics in active galactic nuclei jets). On the contrary, one should treat more complex physical processes involved in a thermal plasma, such as thermal Comptonization, thermal synchrotron radiation, pair creation and annihilation (e.g. \citealt{Svensson82,Coppi90}). For readers' convenience, the fixed parameters of the model are listed in Table \ref{Table:fixed}.

\begin{table}
%\begin{minipage}{17.7cm}
\caption{The fixed parameters for a general microquasar. }
\begin{tabular}{cccc}
\hline\hline
Parameters & Symbol & Value  \\
%&&&$\times 10^4$ &&&$@z_{\rm m}$ &$@z_{\rm m}$ &$@z_{\rm M}$ & $\leq $ &$@z_{\rm M}$ &$\geq $ &&&&\\
\hline
 Black hole mass &  $M_{\rm BH}$  & $10M_{\rm \sun}$   \\
 Surface temperature of companion & $T_{\rm s}$& $6\times10^4\ \rm K$\\
 Radius of companion & $R_{\rm s}$& $10R_{\rm \sun}$\\
 Distance to the Earth & $d$& $2\ \rm kpc$\\
 Orbital radius of binary system & $R_{\rm or}$& $100R_{\rm \sun}$\\
 Accretion rate & $M_{\rm acc}$& $10^{-8}M_{\rm \sun}\ \rm yr^{-1}$\\
 Half-opening angle of the jet & $\varTheta$& $5^{\circ}$\\
 Bulk Lorentz factor of the jet & $\Gamma_{\rm j}$&2\\
 Viewing angle of an observer & $\varphi$&$30^{\circ}$\\
 Thermal electron temperature in the jet & $T_{\rm e}$& $0.001m_{\rm e}c^2$\\
\hline
\end{tabular}
%\\ {\it Notes:} Parameters........
\label{Table:fixed}
%\end{minipage}
\end{table}

\begin{table}
%\begin{minipage}{17.7cm}
\caption{The free parameters of the models.}
\begin{center}
\begin{tabular}{ccccccccc}
\hline\hline
Figures &$\alpha$ & $\zeta$ & $\eta_{\rm jet}$ & $\eta_{\rm e}$ & $\xi$& $r_{\rm 0}$ & $r_{\rm end}$\\
%&&&$\times 10^4$ &&&$@z_{\rm m}$ &$@z_{\rm m}$ &$@z_{\rm M}$ & $\leq $ &$@z_{\rm M}$ &$\geq $ &&&&\\
\hline
Fig. 2  &  2  & 1 &0.1 &0.1 & 10 & $10^4 $&  $10^7$\\
\hline
Fig. 3  &  2  & 1  &0.1  &0.1 & 1     & $10^4 $&  $10^7 $\\
\hline
Fig. 4a    & 2  & 3  &0.1 &0.1 & 1 & $10^4$ &  $10^7$\\
Fig. 4b    & 2  & 3  &0.1 &0.1 & 1 & $10^5$ &  $10^7$\\
Fig. 4c    & 2  & 3  &0.1 &0.1 & 1 & $10^6$ &  $10^7$\\
\hline
Fig. 5a    & 2  & 3  &0.1 &0.1 & 1 & $10^4$ &  $10^5$\\
Fig. 5b    & 2  & 3  &0.1 &0.1 & 1 & $10^4$ &  $10^6$\\
Fig. 5c    & 2  & 3  &0.1 &0.1 & 1 & $10^4$ &  $10^8$\\
\hline
Fig. 6a    & 2  & 10 &0.1 &0.1 & 1 & $10^4 $&  $10^7 $\\
Fig. 6b    & 2  & 1 &0.1 &0.1 & 1 & $10^4 $&  $10^7 $\\
Fig. 6c    & 2  & 0.1 &0.1 &0.1 & 1 & $10^4 $&  $10^7 $\\
\hline
Fig. 7a    & 2  & 1 &0.1 &0.1 & 1 & $10^4 $&  $10^7 $\\
Fig. 7b    & 5/3  & 1 &0.1 &0.1 & 1 & $10^4 $&  $10^7 $\\
Fig. 7c    & 3/2  & 1 &0.1 &0.1 & 1 & $10^4 $&  $10^7 $\\
\hline
Fig. 8    & 2  & 3 &0.1 &0.1 & 1 & $10^4 $&  $10^7 $\\
\hline
Fig. 9a    & 2  & 10 &0.1 &1 & 10 & $10^5 $&  $10^7 $\\
Fig. 9b    & 2  & 10 &0.1 &1 & 1 & $10^5 $&  $10^7 $\\
Fig. 9c    & 2  & 10 &0.1 &1 & 0.1 & $10^5 $&  $10^7 $\\
\hline
\end{tabular}
\end{center}
 {\it Notes.}  Symbol indicating  $\alpha$: spectral index of magnetic turbulence;
$\zeta$: ratio of turbulent magnetic energy density to ordered magnetic one; $\eta_{\rm jet}$: ratio of jet matter power to accretion matter one;  $\eta_{\rm e}$:  ratio of thermal electron power to jet matter one; $\xi$: ratio of magnetic energy density to jet matter one; $r_{\rm 0}$: inner boundary of acceleration zone in units of the gravitational radius $R_{\rm g}$; $r_{\rm end}$: outer boundary of acceleration zone in units of $R_{\rm g}$.
\label{Table:free}
%\end{minipage}
\end{table}

\subsection{Acceleration efficiency of electrons}\label{AEE}
The turbulent magnetized jet consists of a large-scale ordered magnetic field and a small-scale random fluctuation one, in which the latter can help to energize electrons by classical \emph{Fermi} I and II mechanisms. The corresponding acceleration rates are given by equations (\ref{dgdztur}) and (\ref{dgdzsh}), respectively. The cooling rates in a co-spatial acceleration region we consider include adiabatic, synchrotron, its self-Compton scattering, relativistic bremsstrahlung, ICS due to the companion, disc, and background thermal photon fields. The consideration of self-Compton scattering rate is involved in a non-linear process of radiative cooling, in which the synchrotron photon field needs to be calculated in advance via specific distributions of the accelerated electrons obtained in the same jet height. In this section, the injection source $Q(\gamma,r_{\rm 0})$ is assumed to be a Maxwell distribution form.

\begin{figure}
  \begin{center}
  \begin{tabular}{ccc}
\hspace{-0.79cm}
     \includegraphics[width=70mm,height=50mm]{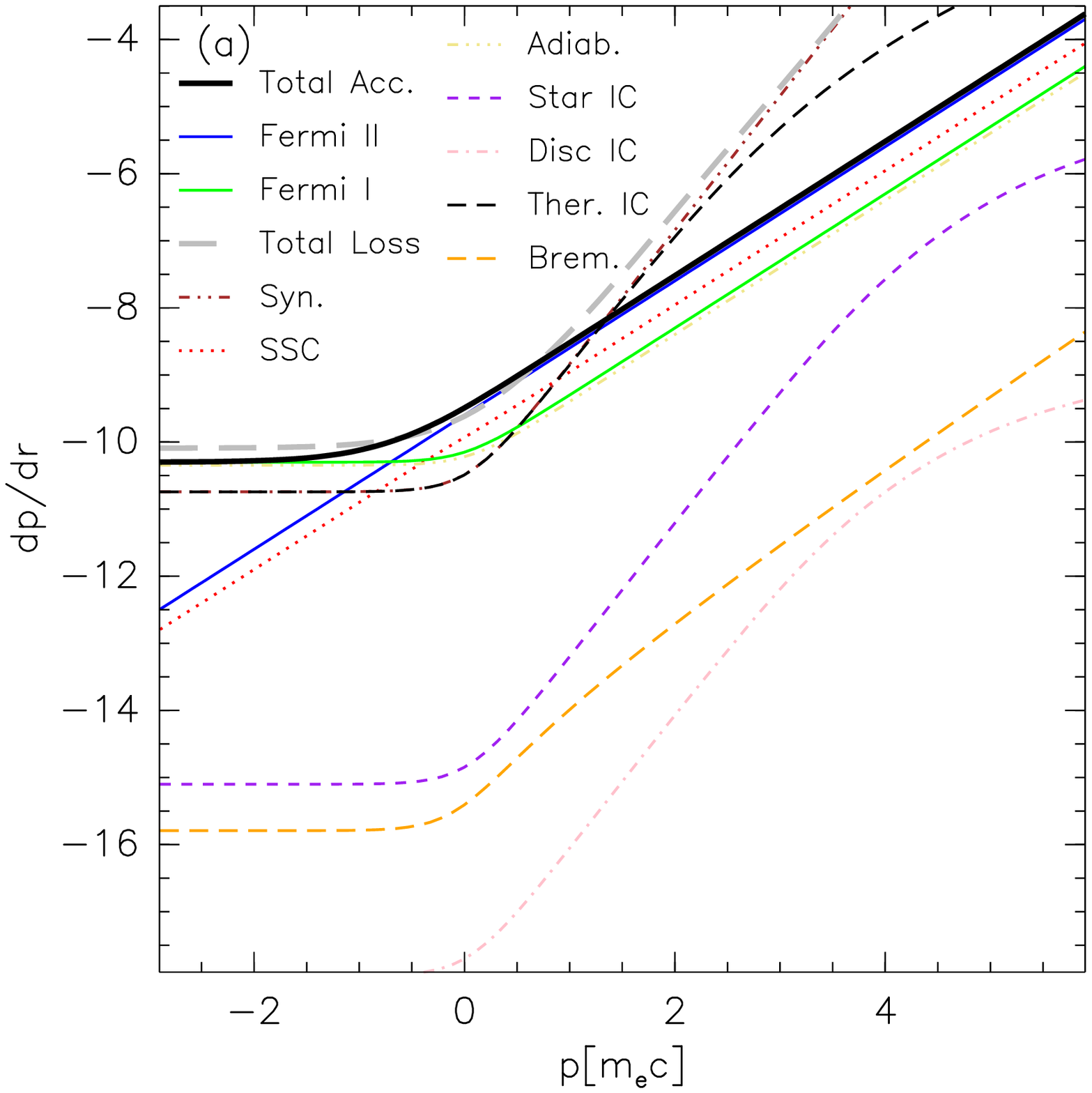} \\%& \ \ \ \ \ \ \ \ \ \ \ \ \ \
\hspace{-0.79cm}
     \includegraphics[width=70mm,height=50mm]{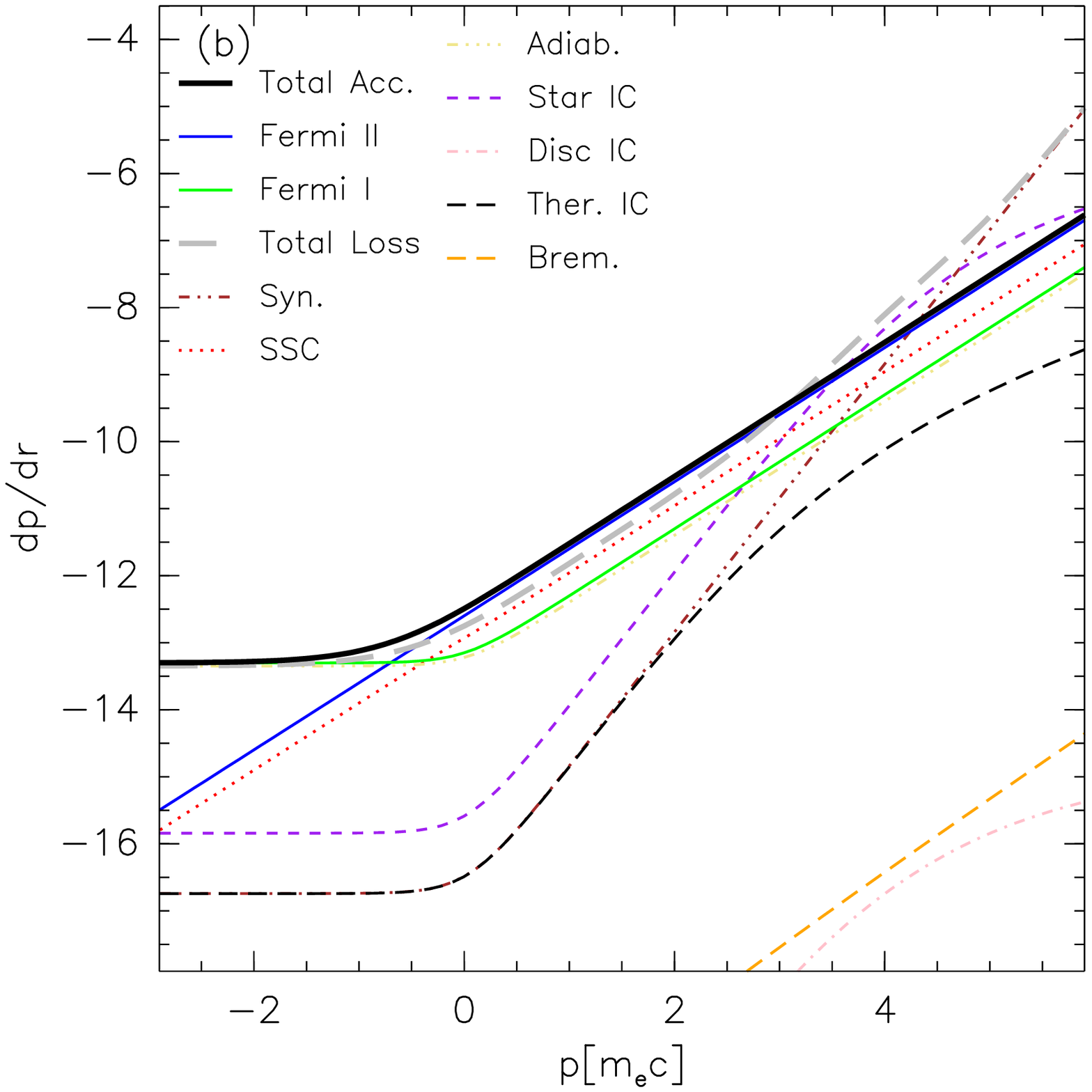}
\end{tabular}
  \end{center}
\caption{Acceleration and cooling rates at the inner ($10^4R_{\rm g}$ for panel a) and outer ($10^7R_{\rm g}$ for panel b) boundaries of an acceleration region located in the jet, as a function of the electron momentum, $p=\beta\gamma$, in units of $m_{\rm e}c$.  The intersection point of the thick solid line (\emph{Total Acc.}) and the thick dashed line (\emph{Total Loss}) corresponds to the equilibrium between total \emph{Fermi} acceleration rates and total loss rates, which determines the maximum energy of accelerated electrons at the boundary location. The used parameters are listed in Tables \ref{Table:fixed} and \ref{Table:free}. Legends indicating \emph{Total Acc.}: the sum of \emph{Fermi} I and II gain rates; \emph{Fermi II}: stochastic acceleration rate; \emph{Fermi I}: diffusive shock acceleration rate; \emph{Total Loss}: the sum of adiabatic and various radiative loss rates; \emph{Syn.}: synchrotron emission loss; \emph{SSC}: synchrotron self-Compton scattering loss; \emph{Adiab.}: adiabatic loss; \emph{Star IC}: ICS loss of stellar photons; \emph{Disc IC}: ICS loss of accretion disc photons;  \emph{Ther. IC}: ICS loss of background thermal photons; \emph{Brem.}: relativistic bremsstrahlung radiative loss. } \label{figs:accloss}
\end{figure}

Fig. \ref{figs:accloss} shows the acceleration and  various cooling curves calculated at the inner ($r_{\rm 0}=10^{4}R_{\rm g}$ for panel a) and outer ($r_{\rm end}=10^{7}R_{\rm g}$ for panel b) boundaries of an acceleration region located in the jet. For parameters used in this figure (see Table \ref{Table:free}), we can find that
the stochastic acceleration rate (\emph{Fermi} II) dominates the diffusive shock acceleration rate (\emph{Fermi} I).  Among these parameters, the extreme value $\xi=10$ denotes that the magnetic pressure is much greater than the jet matter pressure, which implies that the strong magnetic pressure impedes the shock formation. We further test the smaller value $\xi$ (but not shown in this paper), and notice that when the magnetic energy density within the plasma is in sub-equipartition with the jet matter energy density, the shock can be formed and dominates acceleration processes of electrons with decreasing $\xi$, which is consistent with numerical simulations from axisymmetric, magnetically driven relativistic jets (\citealt{Komissarov07}).

As shown in panels (a) and (b), the sum of various cooling rates provides total cooling of electrons, as well as both relativistic bremsstrahlung and ICS cooling rate of accretion disc photons are negligible. The equilibrium between total gain rates and total loss rates can give rise to the maximum momenta of accelerated electrons at different heights of the jet. It can be seen that the maximum momentum of accelerated electrons is $p_{\rm max}\sim 8$ at the initial location of acceleration region (see panel a), and reaches to $p_{\rm max}\sim10^{3}$ at the terminal location of acceleration region (panel b). Provided that a larger value $\zeta$ or a small value $\alpha$ is used, we would obtain a much larger value $p_{\rm max}$, which will be explored in Sects. \ref{TOMF} and \ref{MT}.

\subsection{Electron injection sources}\label{EIS}

\begin{figure*}
  \begin{center}
  \begin{tabular}{ccc}
\hspace{-0.79cm}
     \includegraphics[width=55mm,height=45mm]{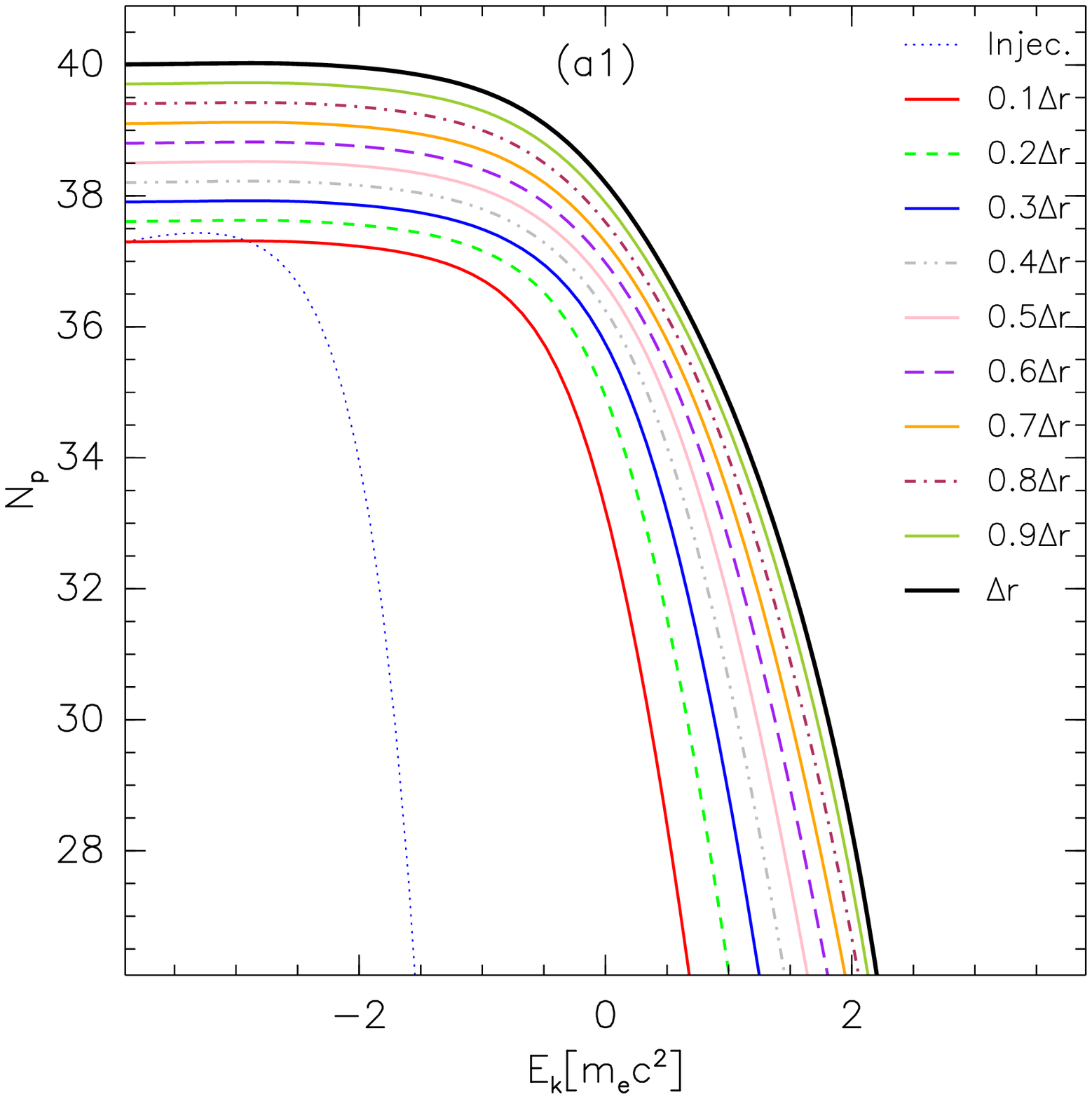}& \ \ \ \ \
\hspace{-0.79cm}
     \includegraphics[width=55mm,height=45mm]{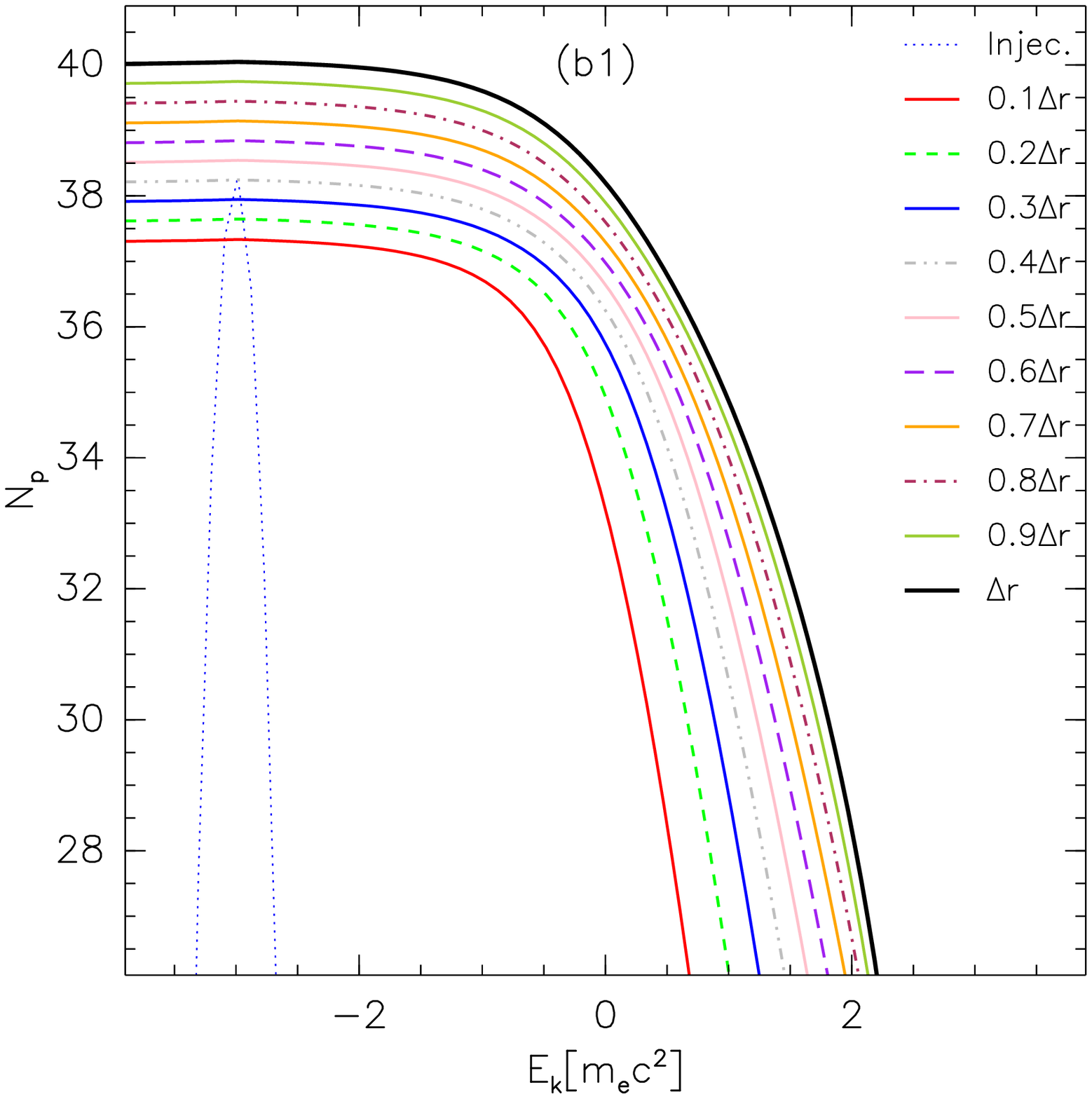}& \ \ \ \ \
\hspace{-0.79cm}
     \includegraphics[width=55mm,height=45mm]{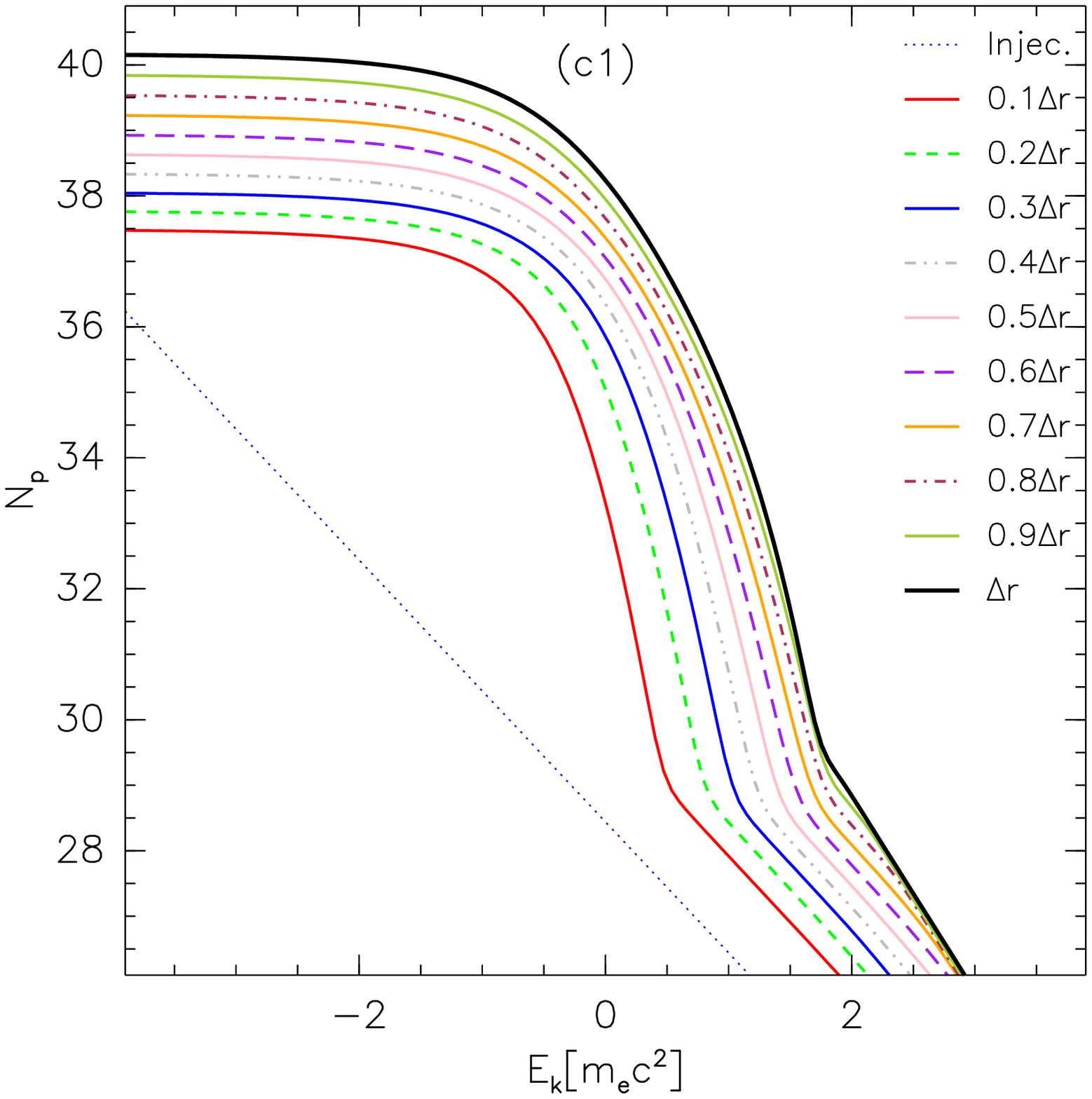}\\
     \hspace{-0.79cm}
     \includegraphics[width=55mm,height=45mm]{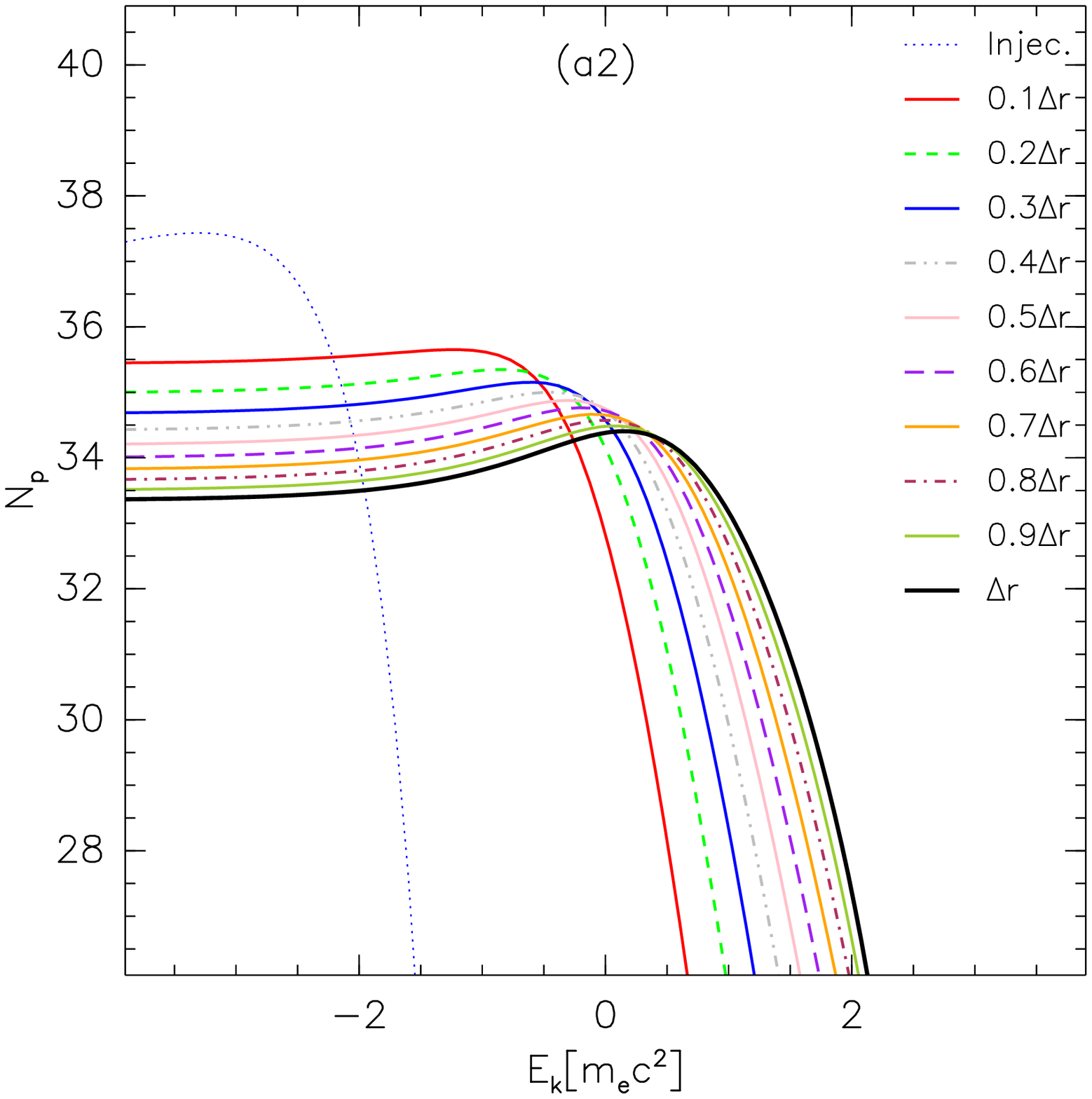}& \ \ \ \ \
\hspace{-0.79cm}
     \includegraphics[width=55mm,height=45mm]{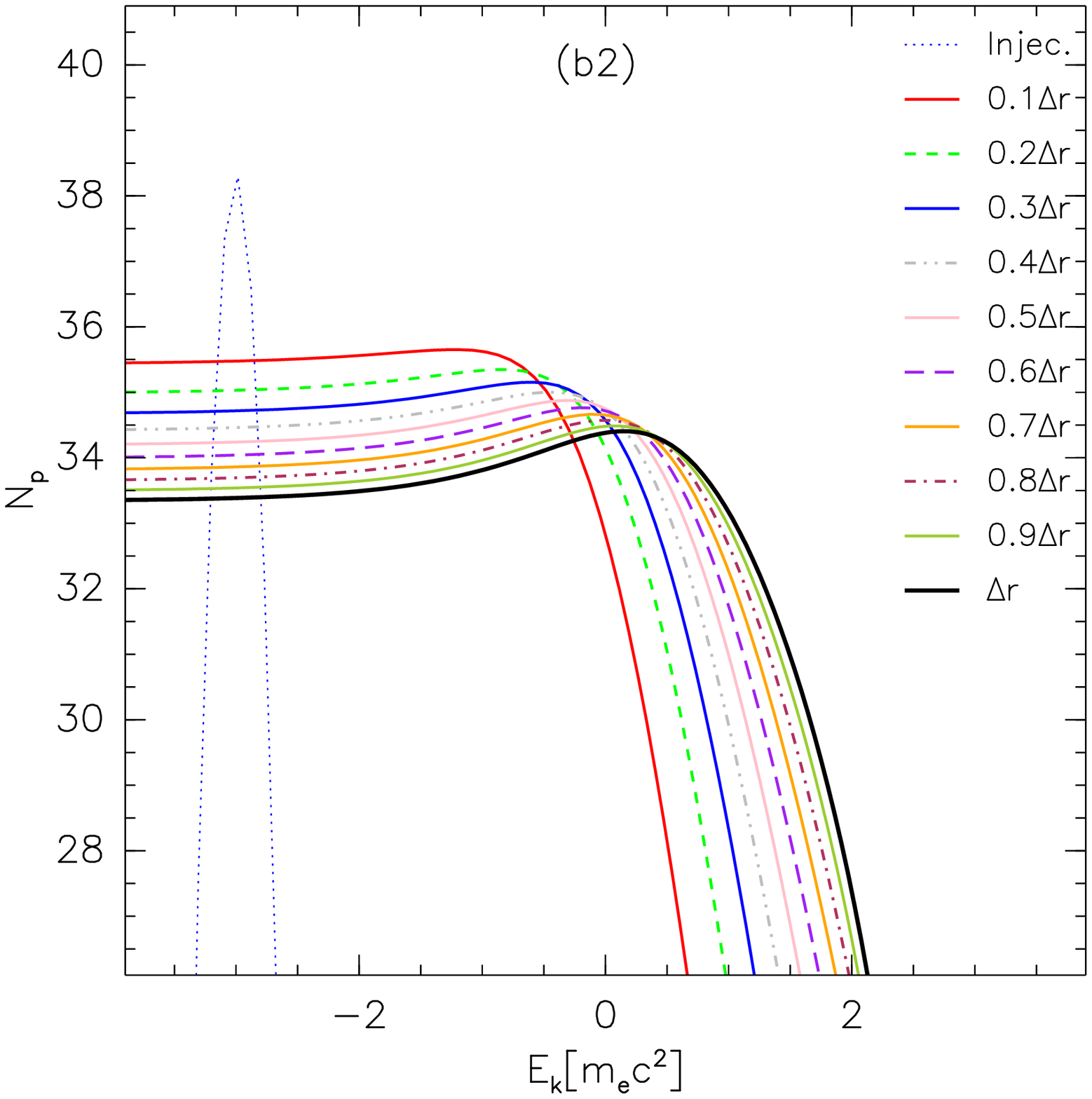}& \ \ \ \ \
\hspace{-0.79cm}
     \includegraphics[width=55mm,height=45mm]{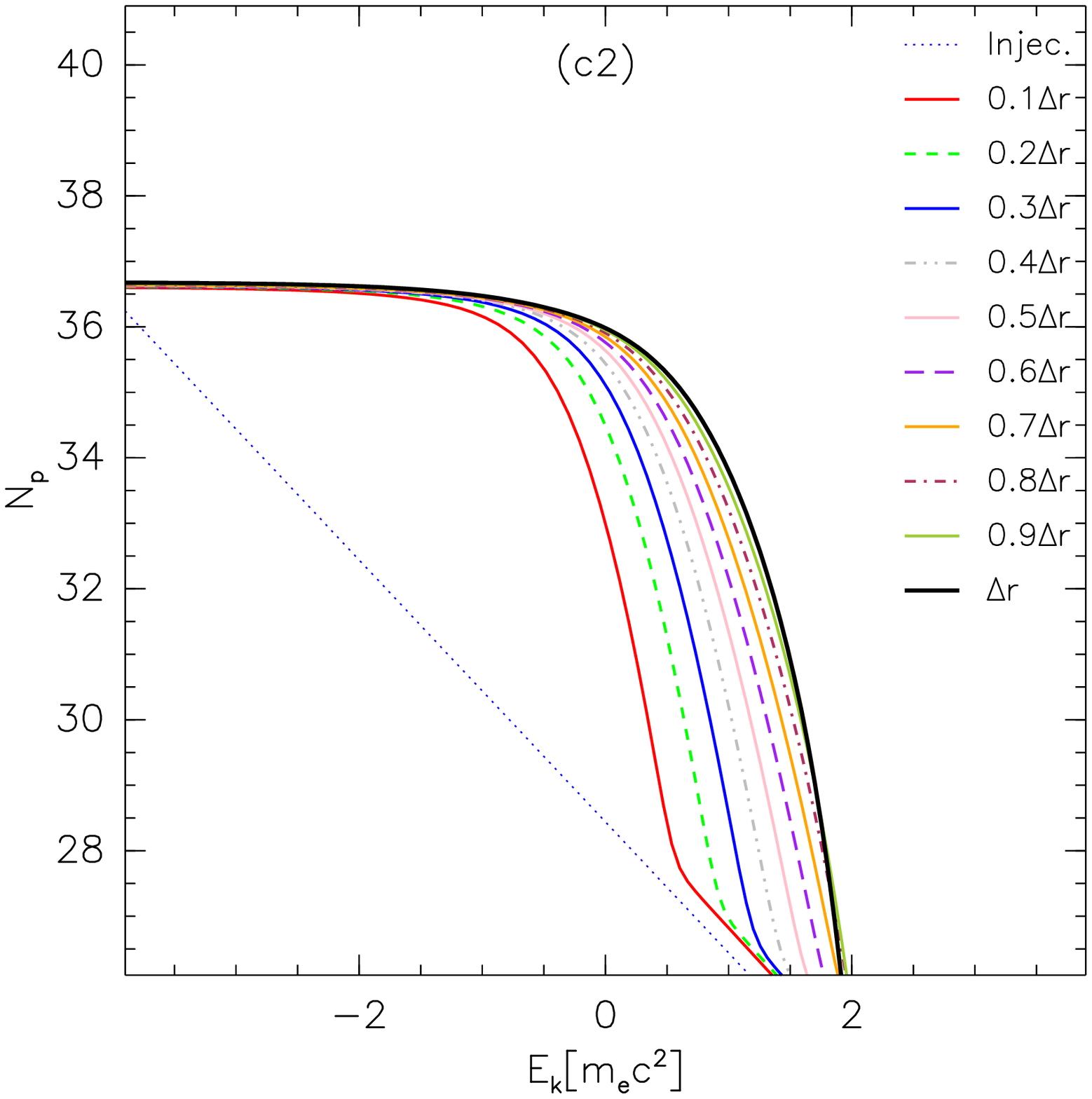}\end{tabular}
  \end{center}
\caption{Spectral energy distributions of accelerated electrons as a function of the electron kinetic energy $E_{\rm k}(=\gamma-1)$ in units of $m_{\rm e}c^2$, with continuous injection source of $Q(p,r)$ (upper panels a1--c1) along the jet and $\delta(r_{\rm 0})$ function injection of $Q(p,r)$ (lower panels a2--c2).  Panels (a1) and (a2) are for a Maxwell distribution of accelerated electrons with the equilibrium temperature of $T_{\rm e}=0.001$ in units of $m_{\rm e}c^2$; panels (b1) and (b2) for a Gaussian distribution with the expectation of distribution of $\mu=T_{\rm e}$ and the standard derivation of $\sigma=0.1$; as well as panels (c1) and (c2) for a non-thermal power-law form with the spectral index of 2. Electron spectral distributions are plotted in each panel in a logarithmic interval of $0.1\Delta r$ between the curves, with $\Delta r=3$. The parameters we used are listed in Tables \ref{Table:fixed} and \ref{Table:free}.}  \label{figs:deltayesno}
\end{figure*}

We here explore electron injection sources with three distribution forms: Maxwell, Gaussian and power-law distributions. On the basis of these distributions, we take two spatial injection ways into account, that is, one way is that electrons are continuously injected along the whole acceleration region, and the other way is that they are injected only at the inner boundary of the acceleration region. The numerical results are presented in Fig. \ref{figs:deltayesno}, the upper panels of which correspond to continuous injection source of $Q(p,r$), and the lower panels of which to $\delta(r_{\rm 0})$ function injection of $Q(p,r)$. The model parameters for different cases are listed in Tables \ref{Table:fixed} and \ref{Table:free}. Electron spectral energy distributions are plotted in each panel in a logarithmic interval of $0.1\Delta r$ between the curves. It is noted that logarithmic range of the acceleration region is $\Delta r=3$ and the logarithmic step used in our numerical simulations is $\Delta r/200=0.015$.

In panels (a1) and (a2) of Fig. \ref{figs:deltayesno}, we plot the result of electron accelerations with an initial Maxwell distribution, whose equilibrium temperature is set as $T_{\rm e}=0.001$ in units of $m_{\rm e}c^2$. For the continuous injection case (Fig. \ref{figs:deltayesno}a1), the injected electrons are gradually accelerated up to high energies with increasing the height of the jet $r$, and the resulting spectra form a non-thermal tail but have a Maxwell-like distribution shape (in this paper, the accelerated electrons are referred to as non-thermal electrons since they are at the non-thermal equilibrium state). It is very obvious that electron fluxes (e.g. the number of electrons per unit energy) increase with $r$ due to a continuous injection of thermal electrons.  As for $\delta(r_{\rm 0})$ function injection case (Fig. \ref{figs:deltayesno}a2), the injected thermal electrons are gradually energized beyond the thermal background into more high-energy regime. The spectral distributions of the accelerated electrons also demonstrate a non-thermal tail. Due to the lack of a continuous injection of thermal electrons, the number of electrons accelerated in low-energy regime decreases gradually, because the accelerated low-energy electrons would be re-energized into more high-energy regime.

The panels (b1) and (b2) of Fig. \ref{figs:deltayesno} present the results for a Gaussian injection with the expectation of distribution of $\mu=0.001$ and the standard derivation of $\sigma=0.1$. As shown in the panels, the resulting spectral distributions are the same as those in the case of the Maxwell injection (see also panels a1 and a2). As a result, the background low-energy thermal electrons with Maxwell or Gaussian distribution can produce the same spectral shape of the relativistic electrons.

In general, spectral distributions resulting from a power-law injection (with the spectral index of 2) also present a Maxwell-like distribution at low-energy region but show a transition to form a power-law shape with the index close to 2 at the front part of acceleration. Along with the advancement of acceleration process, the spectra at the high-energy region become steeper since the radiative cooling impedes the further acceleration of electrons, and these accelerated relativistic electrons begin to cool down rapidly (see panels c1 and c2 in Fig. \ref{figs:deltayesno} ). In brief, the acceleration processes we have studied can produce not only an increase of electron energies, but also result in a broadening of their spectral energy distributions.

\subsection{Different acceleration regions}\label{DAR}
\begin{figure*}
  \begin{center}
  \begin{tabular}{cccc}
\hspace{-0.79cm}
     \includegraphics[width=55mm,height=50mm]{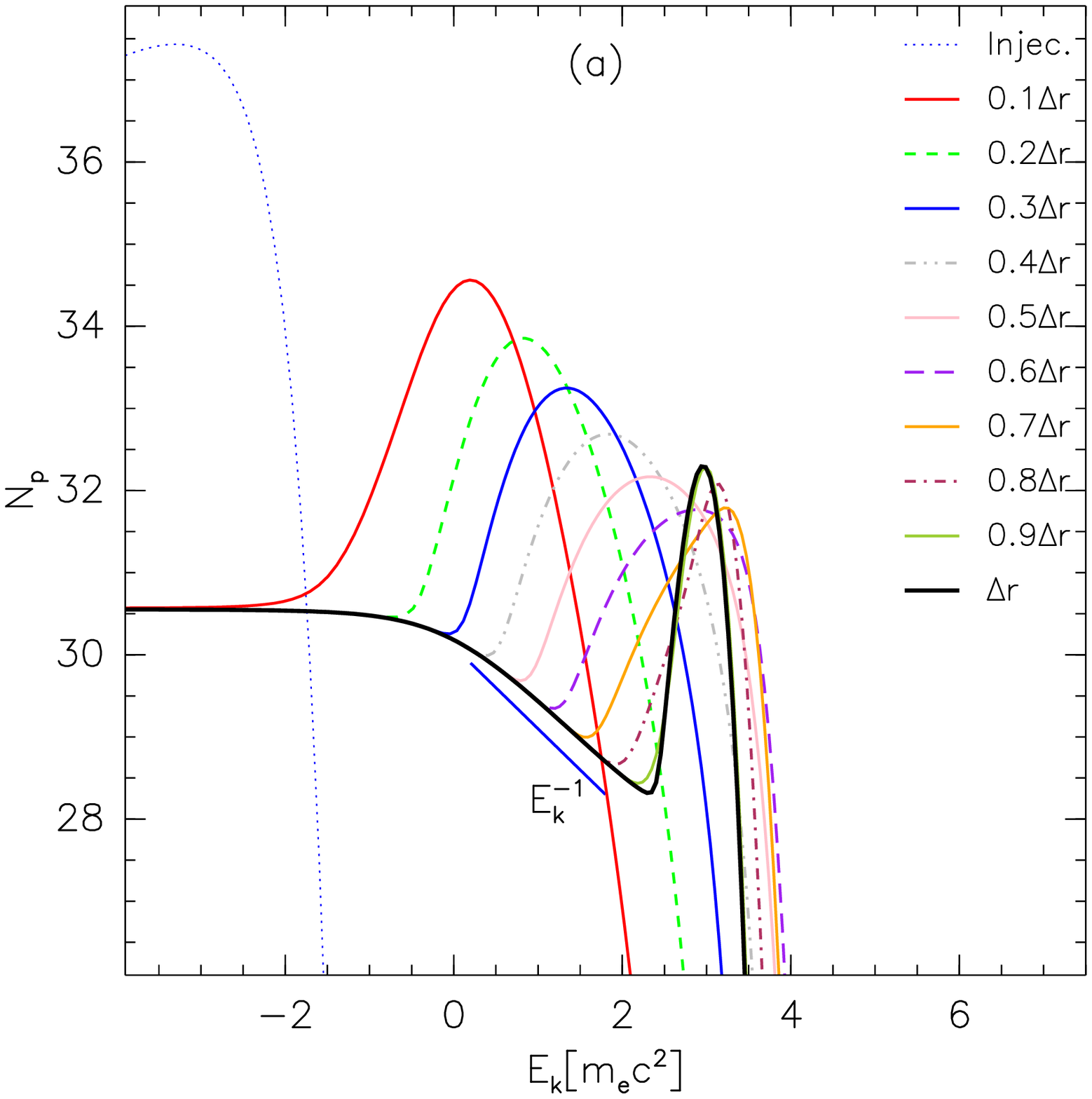}& \ \ \ \ \
\hspace{-0.79cm}
     \includegraphics[width=55mm,height=50mm]{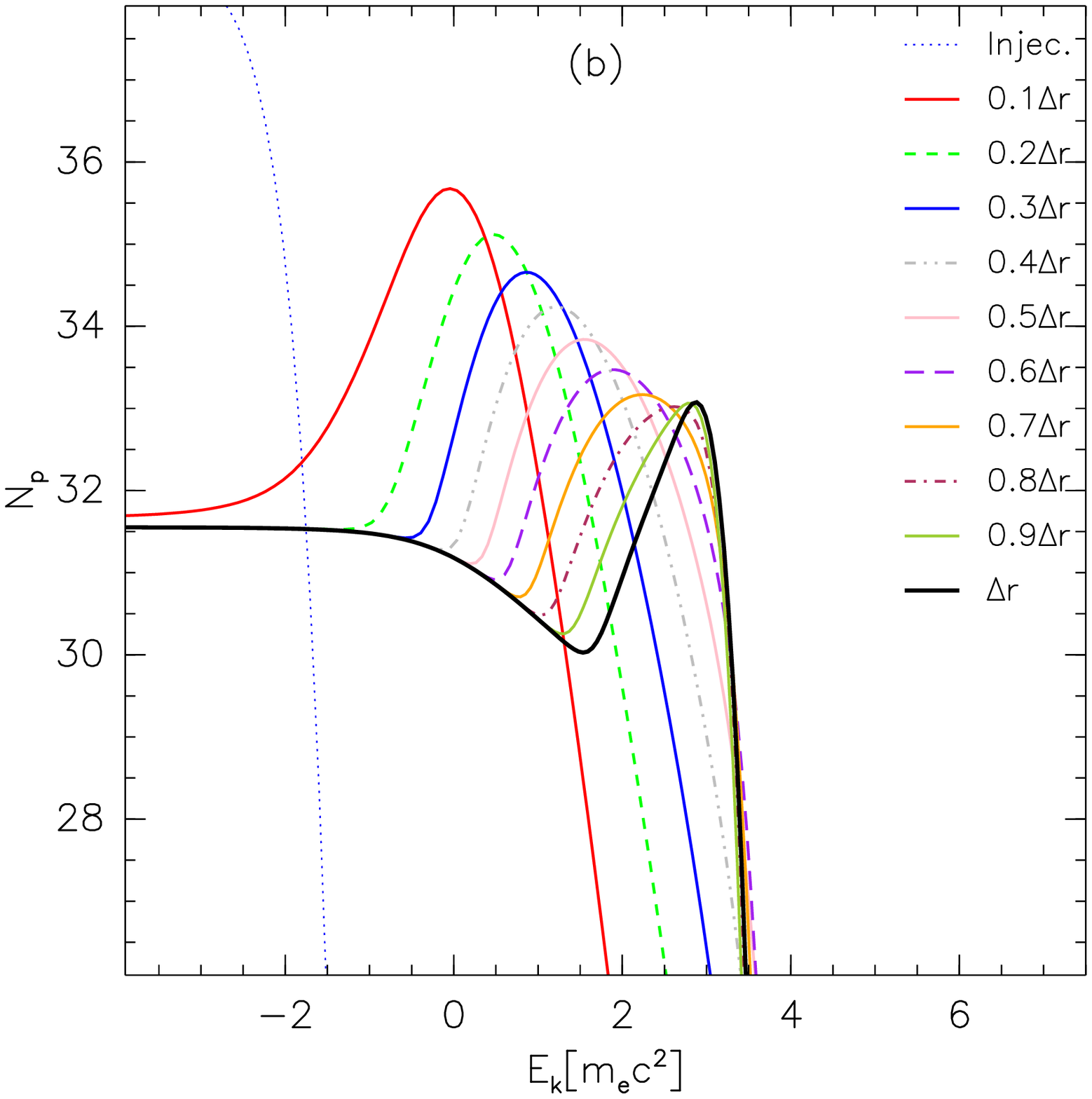}& \ \ \ \ \
\hspace{-0.79cm}
     \includegraphics[width=55mm,height=50mm]{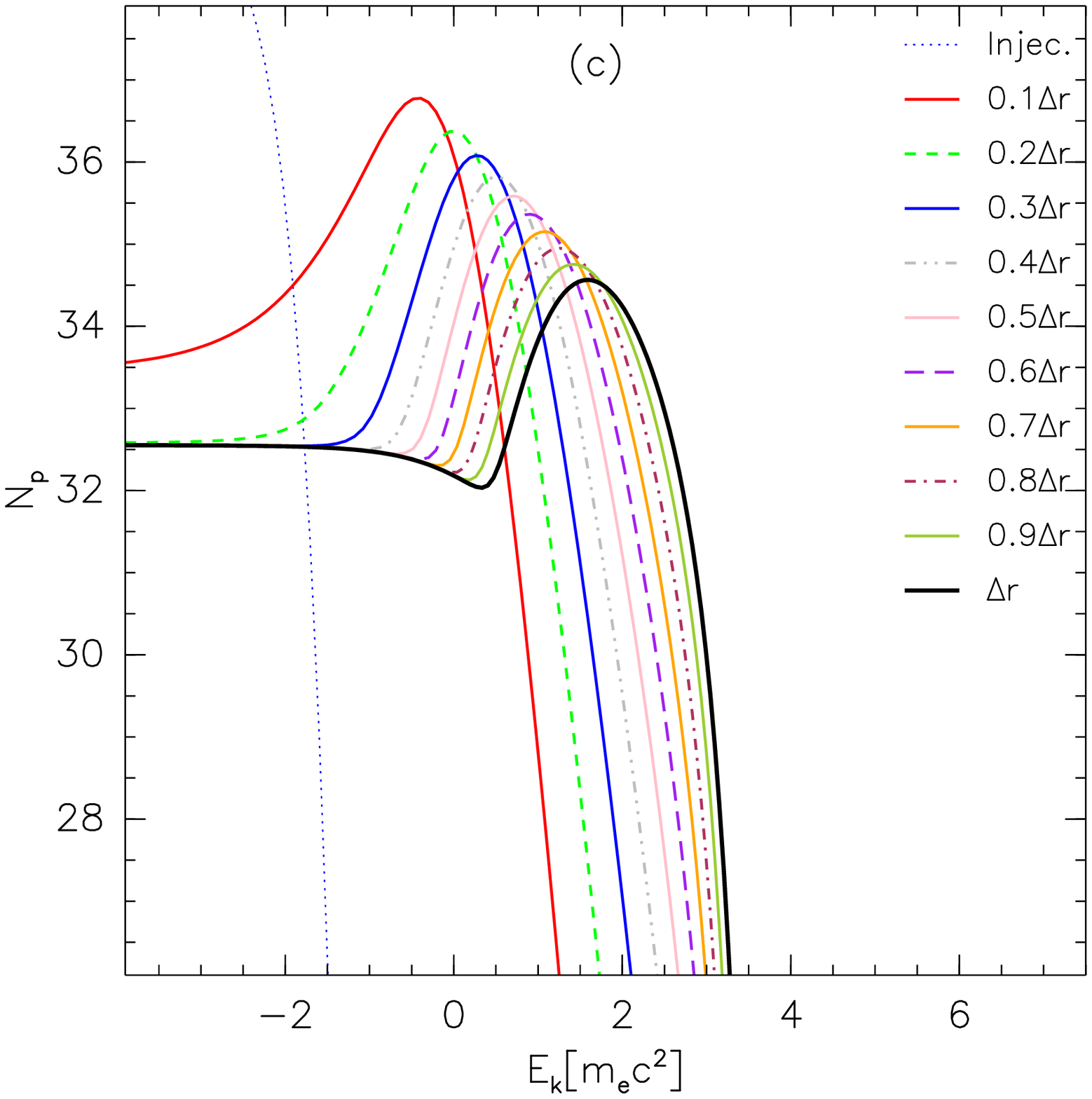}
\end{tabular}
  \end{center}
\caption{Electron spectral distributions for different inner boundaries of acceleration region: $r_{\rm 0}=10^4R_{\rm g}$ for panel (a),  $10^5R_{\rm g}$ for panel (b) and  $10^6R_{\rm g}$ for panel (c). The $\delta(r_{\rm 0})$ function injection source at $r_{\rm 0}$ is considered and the outer boundary of the acceleration region is set as $r_{\rm end}=10^7R_{\rm g}$, which corresponds to the logarithmic range of $\Delta r=3$ for panel (a), $\Delta r=2$ for panel (b) and $\Delta r=1$ for panel (c),  with the logarithmic interval of $0.1\Delta r$, respectively. The model parameters are listed in Tables \ref{Table:fixed} and \ref{Table:free}.
}  \label{figs:AcceRegion}
\end{figure*}

The effects of different acceleration regions (varying inner and outer boundary conditions) on the accelerated electrons' spectra are explored in this section. We first fix the outer boundary of the acceleration region as $r_{\rm end}=10^7R_{\rm g}$, which is located outside the binary system scale $R_{\rm or}$. The inner boundary of the acceleration region are set as $r_{\rm 0}=10^4R_{\rm g}$ for panel (a) of Fig. \ref{figs:AcceRegion}, $10^5R_{\rm g}$ for panel (b) and $10^6R_{\rm g}$ for panel (c). Furthermore, the injection source is assumed to be a thermal Maxwell distribution, with $\delta(r_{\rm 0})$ function injection form in the aspect of space. The relevant parameters used are listed in Tabels \ref{Table:fixed} and \ref{Table:free}.

As shown in Fig. \ref{figs:AcceRegion}(a), background electrons are energized into high energies beyond their thermal distribution to form a new Gaussian-like tail distribution, the peak of which is determined by equilibrium between acceleration rates and various cooling rates, where most of electrons pile up around it. With increasing the height of the jet $r$, the peak shifts to more high energies but decreases in fluxes. However, when approaching to the outer boundary of the acceleration zone, the full width at half-maximum of Gaussian-like distributions becomes small and the number of accelerated electrons increases due to strong radiative losses suppressing electrons up to more high energies. Below the peak energy, the spectra present a power-law form (with index of 1) due to a decrease of the number of low-energy electrons.

When the value of the inner boundary of acceleration zone, $r_{\rm 0}$, is increased, the power-law component gradually fades away, and the peak of distributions, which has a larger full width at half-maximum of Gaussian-like distributions, shifts to low-energy regime  (see panels b and c). It can be seen that the peak energy of the accelerated electron population would decrease with increasing the value of $r_{\rm 0}$, but increase in their fluxes. In addition, the maximum energy reached in the acceleration process also slightly decrease due to a relatively low level of magnetic turbulence. On the contrary, we also test smaller values of $r_{\rm 0}$, such as $r_{\rm 0}=10^2R_{\rm g}$ and $10^3R_{\rm g}$, and find that peaks shift to much lower-energy regime during the whole acceleration process, because of both much stronger radiative cooling and more inefficient acceleration. The number of electrons increases with increasing the inner boundary of the acceleration region, because electron injections are related to becoming larger spatial location $r_{\rm 0}$ of the jet. Consequently, an efficient acceleration process of relativistic particles should be produced at a distance that is greater than $10^{3}R_{\rm g}$ for the common parameter settings.

\begin{figure*}
  \begin{center}
  \begin{tabular}{cccc}
\hspace{-0.79cm}
     \includegraphics[width=55mm,height=50mm]{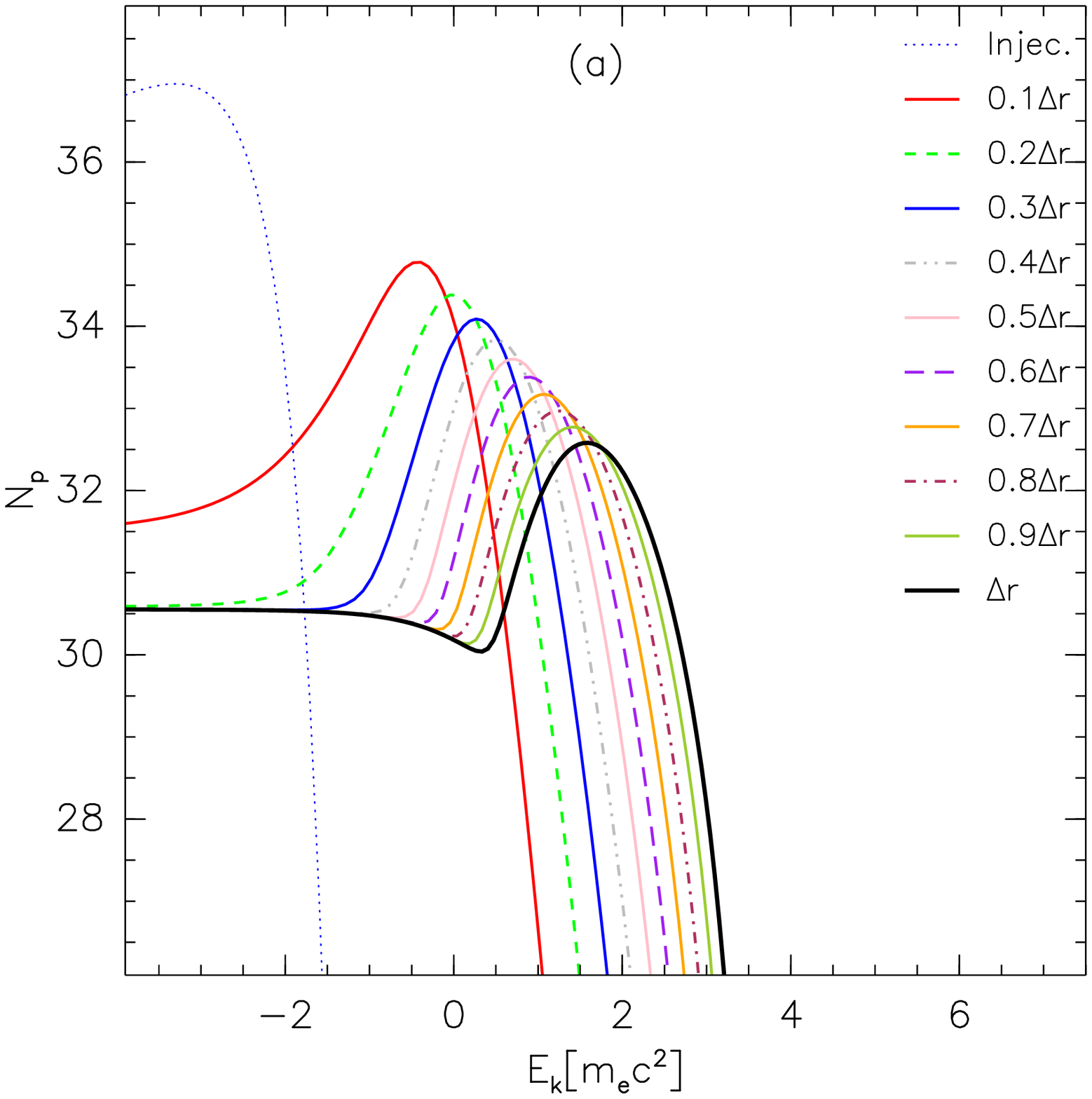}& \ \ \ \ \
\hspace{-0.79cm}
     \includegraphics[width=55mm,height=50mm]{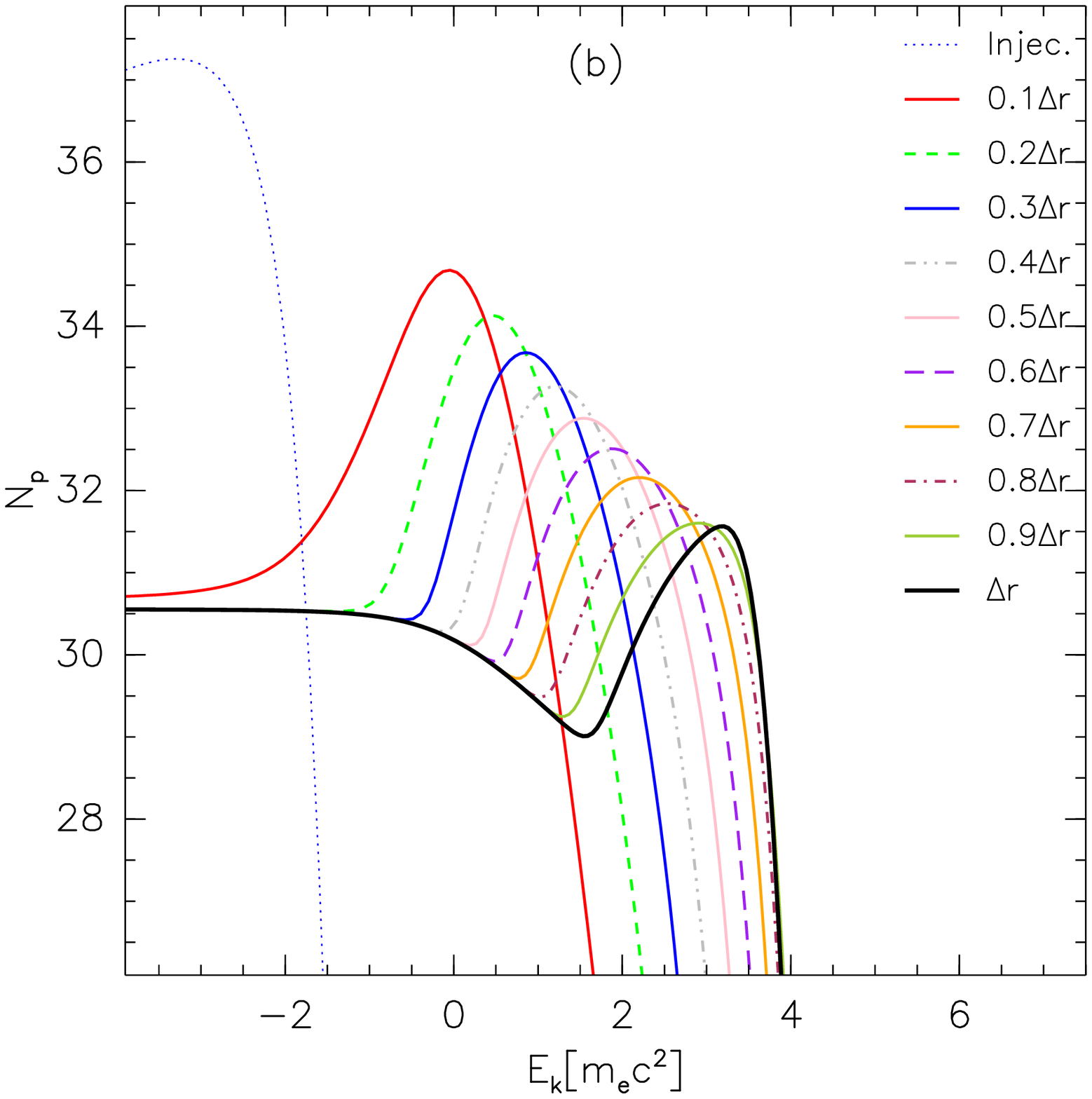}& \ \ \ \ \
\hspace{-0.79cm}
     \includegraphics[width=55mm,height=50mm]{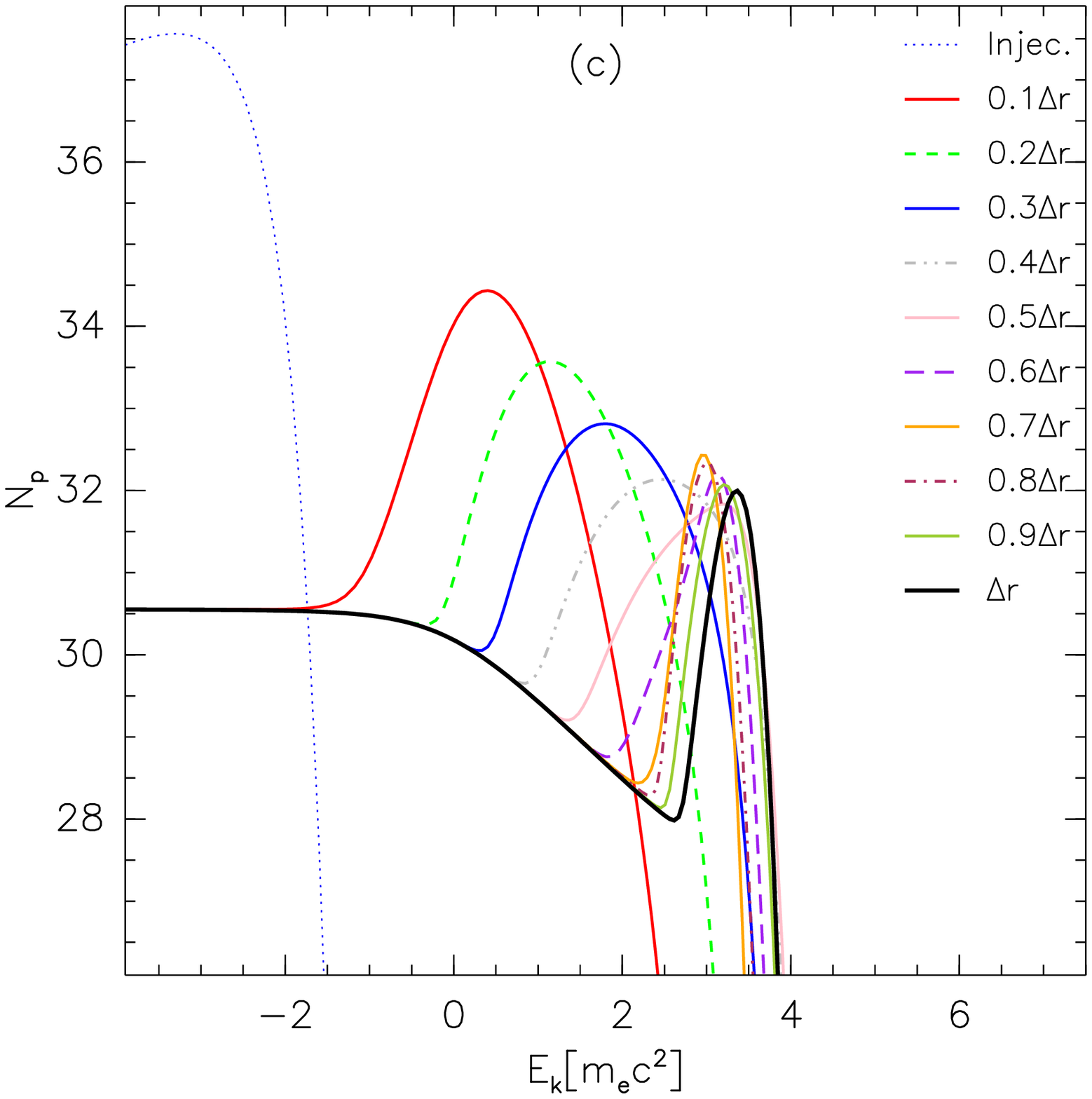}
\end{tabular}
  \end{center}
\caption{Electron spectral distributions for different outer boundaries of acceleration region: $r_{\rm end}=10^5R_{\rm g}$ for panel (a),  $10^6R_{\rm g}$ for panel (b) and  $10^8R_{\rm g}$ for panel (c). The $\delta(r_{\rm 0})$ function injection source at the inner boundary of the acceleration region is set as $r_{\rm 0}=10^4R_{\rm g}$, which corresponds to the logarithmic range of $\Delta r=1$ for panel (a), $\Delta r=2$ for panel (b) and $\Delta r=4$ for panel (c),  with the logarithmic interval of $0.1\Delta r$, respectively. The model parameters are listed in Tables \ref{Table:fixed} and \ref{Table:free}.
}  \label{figs:AcceRegion1}
\end{figure*}

We then fix the inner boundary value as $r_{\rm 0}=10^4R_{\rm g}$, and change outer boundary values: $r_{\rm end}=10^5R_{\rm g}$ for panel (a) of Fig. \ref{figs:AcceRegion1},  $10^6R_{\rm g}$ for panel (b) and $10^8R_{\rm g}$ for panel (c), to study the behavior of relativistic electron spectra (see Fig. \ref{figs:AcceRegion1}). As seen in Fig. \ref{figs:AcceRegion1}, the general behavior of spectral distributions is similar to those of Fig. \ref{figs:AcceRegion}. The difference is that the maximum energy of electrons increase with increasing the value $r_{\rm end}$, and then saturate when reaching the location of $\sim 10^6R_{\rm g}$ (see Fig. \ref{figs:AcceRegion}a and Fig. \ref{figs:AcceRegion1}). In this paper, we therefore fix the outer boundary of the acceleration region at $r_{\rm end}=10^7R_{\rm g}$ beyond the scale of the orbital radius of binary system for simplicity.

\subsection{Turbulent vs. ordered magnetic fields}\label{TOMF}
It is well known that magnetic turbulence is very ubiquitous in astrophysics and plays an important role in some key astrophysical processes, such as acceleration and propagation of cosmic rays, star formation, heat conduction, magnetic reconnection and accretion processes. However, in the X-ray binary jet environment, there is very little work involving a turbulent magnetic field configuration (but see \citealt{Zhang17}). This work is studying the case that magnetized jets mixing turbulent and ordered magnetic fields induce stochastic diffusion and shock collision interaction, which can result in the acceleration of electrons. We here study how relative strength between the turbulent magnetic field and ordered one impacts the acceleration of electrons.

\begin{figure*}
  \begin{center}
  \begin{tabular}{cccc}
\hspace{-0.79cm}
     \includegraphics[width=55mm,height=50mm]{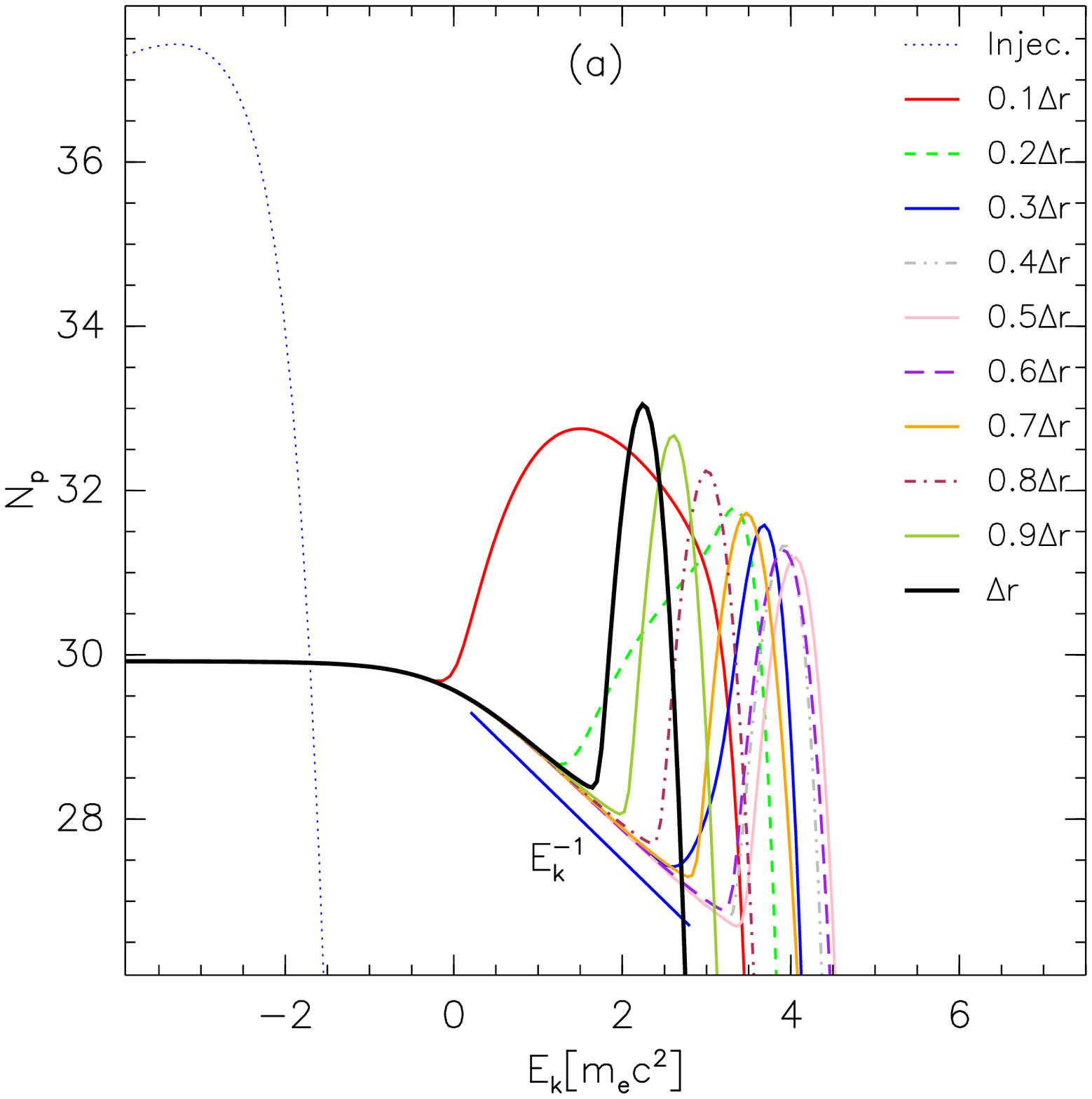}& \ \ \ \ \
\hspace{-0.79cm}
     \includegraphics[width=55mm,height=50mm]{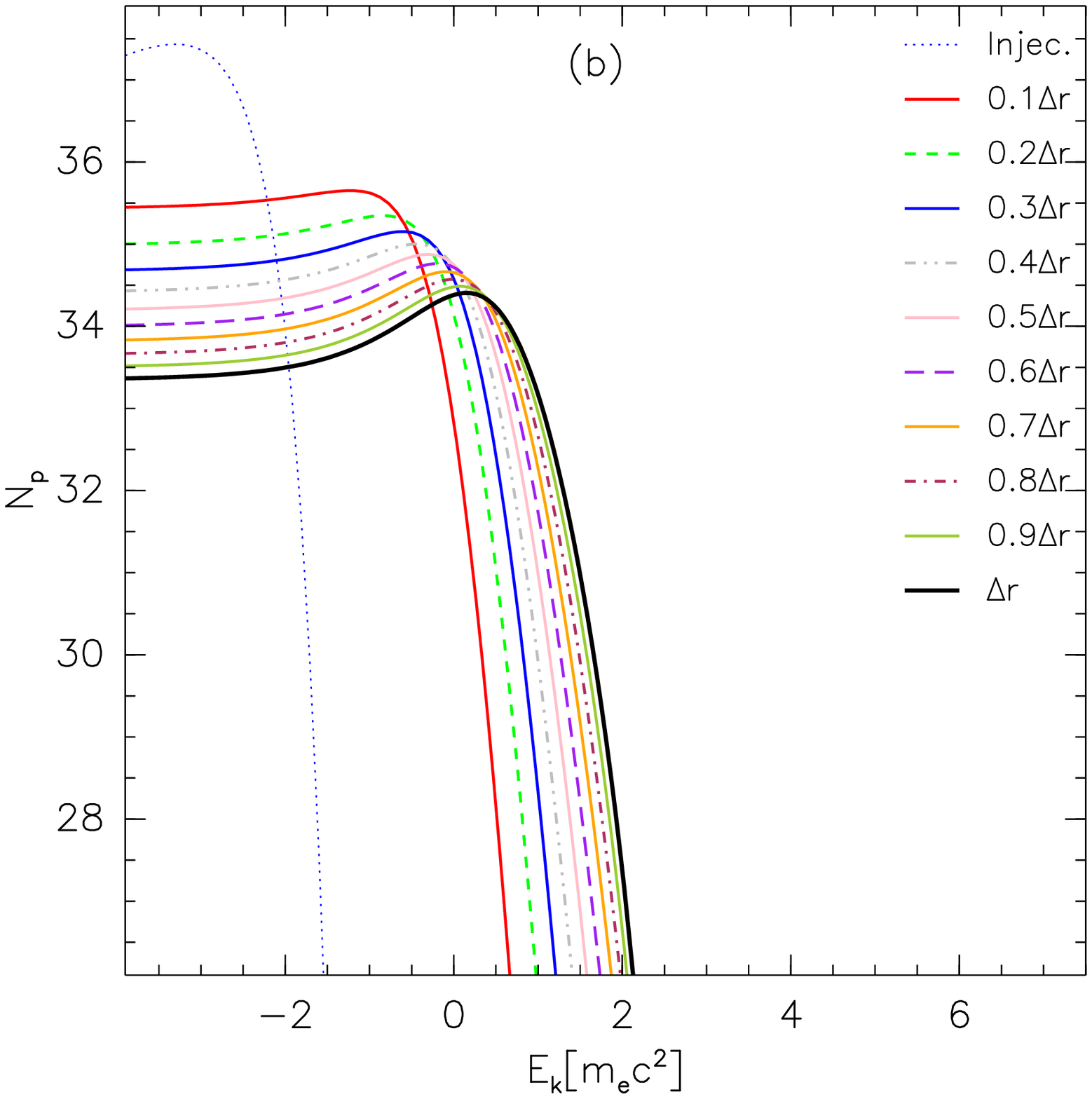}& \ \ \ \ \
\hspace{-0.79cm}
     \includegraphics[width=55mm,height=50mm]{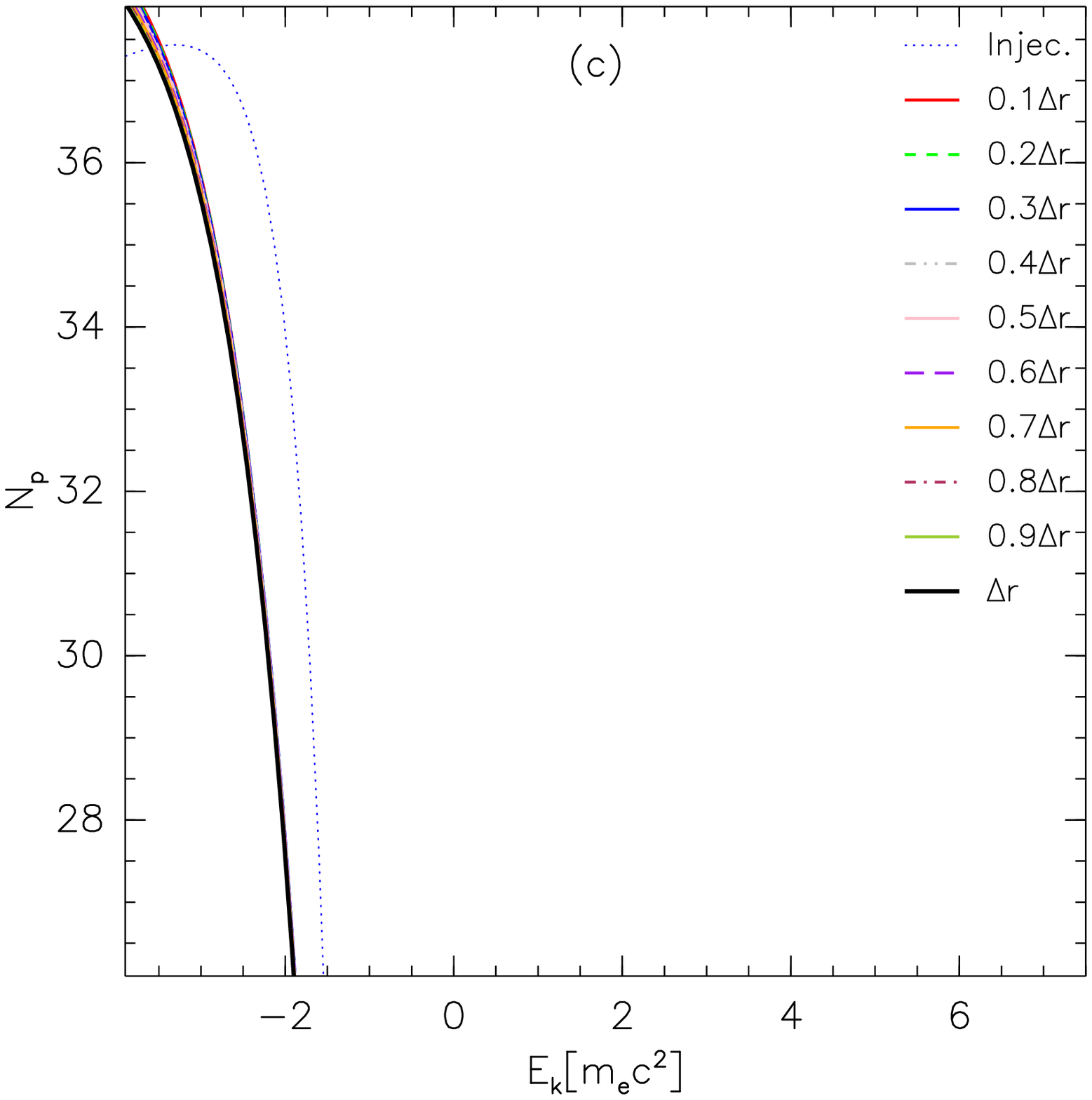}
\end{tabular}
  \end{center}
\caption{Electron spectral distributions for different ratios of the turbulent magnetic field energy density to the ordered one: $\zeta=\delta B^2/B^2=10$ for panel (a), $\zeta=1$ for panel (b), and $\zeta=0.1$ for panel (c). The $\delta(r_{\rm 0})$ function injection source at $r_{\rm 0}$ is adopted in this scenario. Spectral distributions are plotted in a logarithmic interval of $0.1\Delta r$ between the curves with $\Delta r=3$. The adopted parameters are listed in Tables \ref{Table:fixed} and \ref{Table:free}.}  \label{figs:Magratio}
\end{figure*}

The resultant spectral energy distributions are presented in Fig. \ref{figs:Magratio} for the ratio of turbulent field energy density to ordered one:  $\zeta=\delta B^2/B^2=10.0$ for panel (a), $\zeta=1.0$ for panel (b), and $\zeta=0.1$ for panel (c). The background electrons with thermal Maxwell distribution are injected only at $10^4R_{\rm g}$, but the acceleration region is extended up to $10^7R_{\rm g}$. The model parameters we
used are listed in Tables \ref{Table:fixed} and \ref{Table:free}. Panel (a) shows the case of a large ratio, i.e. a dominant magnetic turbulence case. The resulting spectra are broadened beyond background thermal distribution at the front part of acceleration zone, and narrowed down to form Gaussian-like shape at the late part due to radiative losses of high-energy electrons.

The equipartition case is explored in panel (b), which is the same as that of Fig. \ref{figs:deltayesno}a2 and is explained in Sect. \ref{EIS}. The numerical results for the case of the dominant ordered magnetic fields are shown in panel (c), in which spectra shift towards much lower energies than the given initial injection spectrum, due to both radiative cooling and the lack of any acceleration. It follows that it is not possible for a weak turbulence to accelerate low-energy electrons to high energies by \emph{Fermi} I or II process. In other words, the magnetized turbulent environment plays a critical role in accelerating particles and even producing broadband electromagnetic radiation.

\subsection{Magnetic Turbulence}\label{MT}
Using an analytical, especially, numerical simulation method to investigate MHD turbulence and its implications has achieved a significant
success (e.g. \citealt{Goldreich95,Cho02,Lazarian99,Kowal09, Beresnyak15}). In particular, \cite{Lazarian16} proposed a new technique based on observations to reveal the properties of MHD turbulence, which has been successfully tested by \cite{Zhang16} and \cite{Lee16} and can be applied to an astrophysical research. However, we notice that many outstanding questions such as the spectral slope of MHD turbulence are needed to further study.

\begin{figure*}
  \begin{center}
  \begin{tabular}{cccc}
\hspace{-0.79cm}
     \includegraphics[width=55mm,height=50mm]{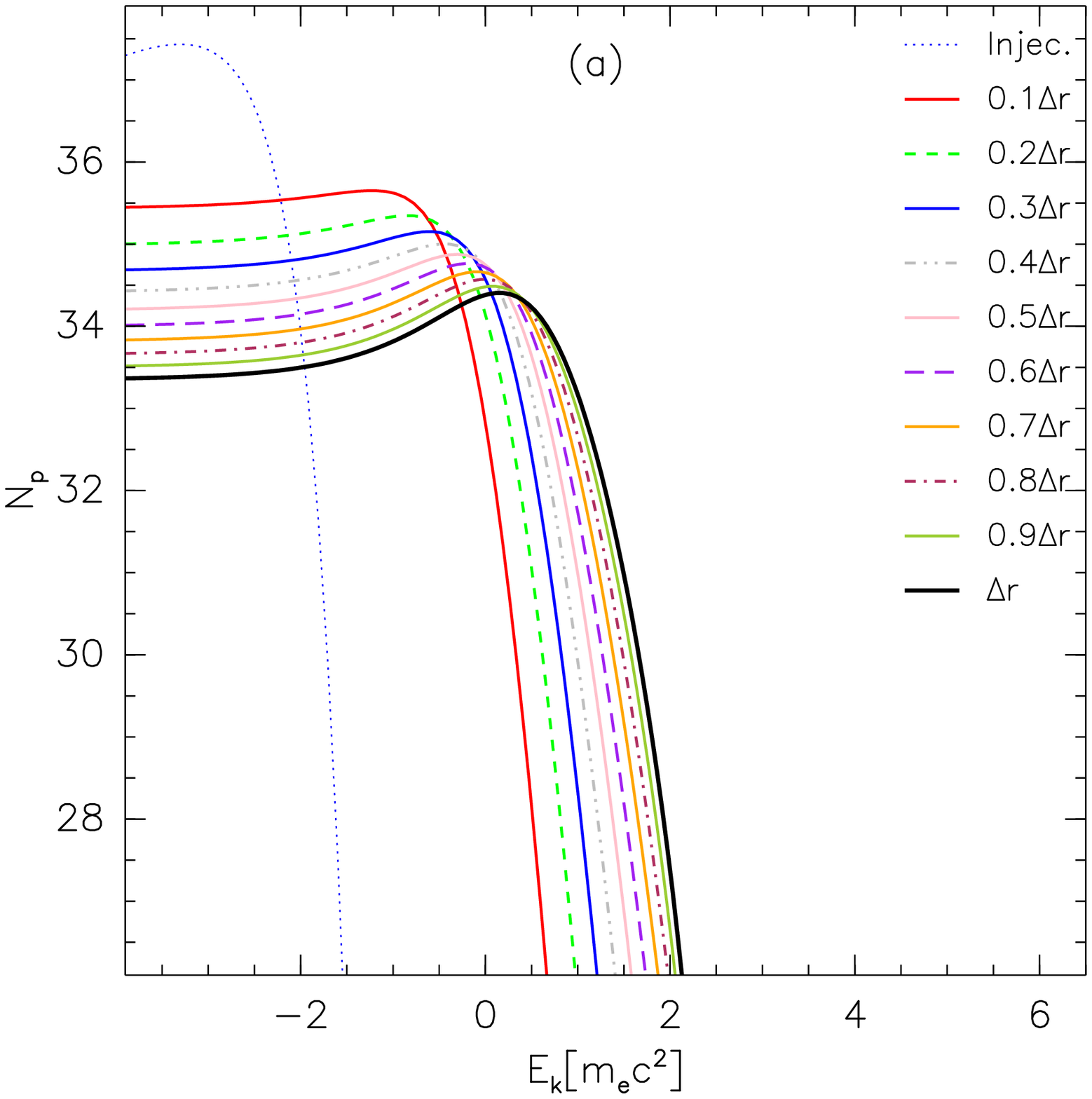}& \ \ \ \ \
\hspace{-0.79cm}
     \includegraphics[width=55mm,height=50mm]{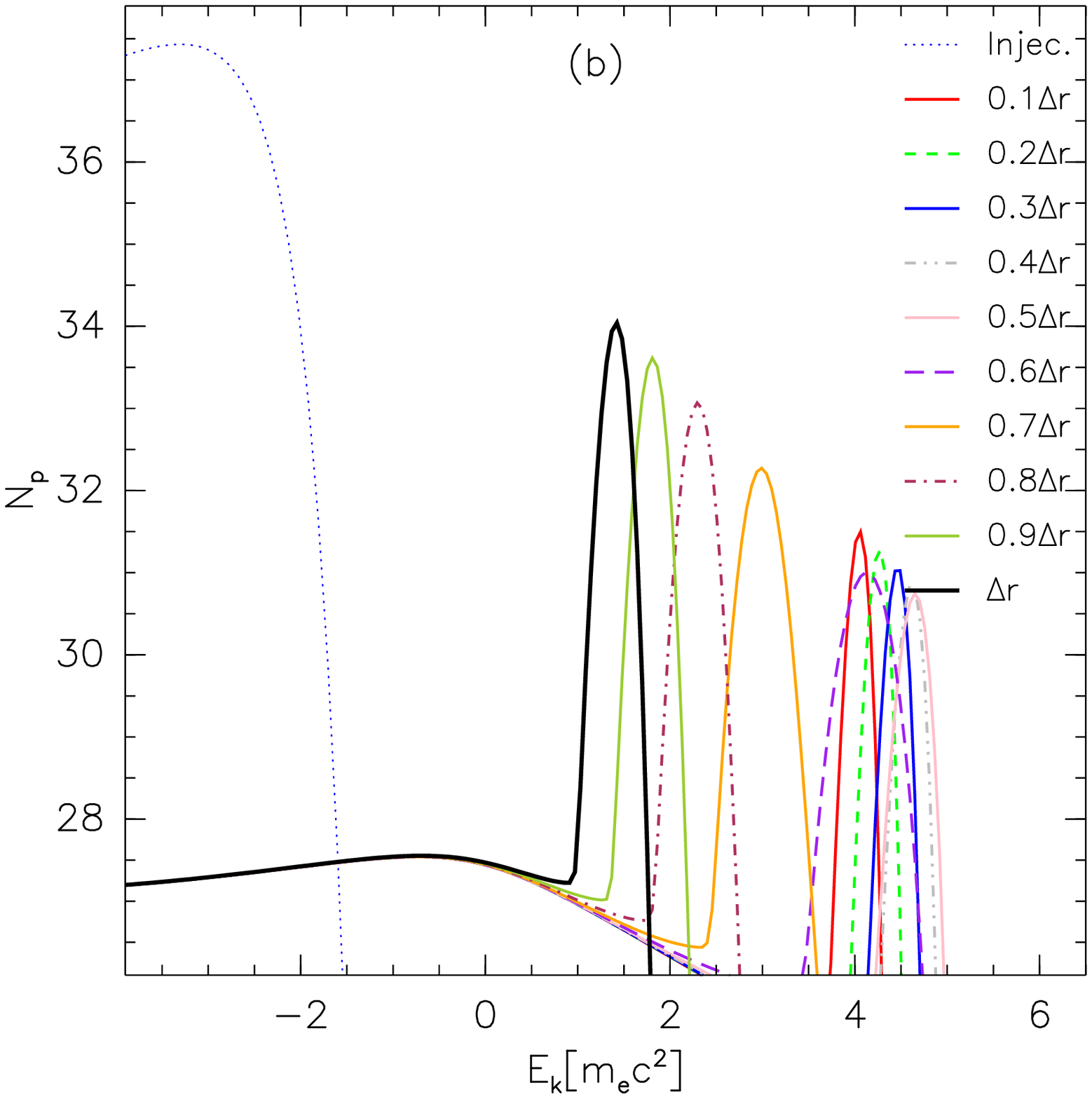}& \ \ \ \ \
\hspace{-0.79cm}
     \includegraphics[width=55mm,height=50mm]{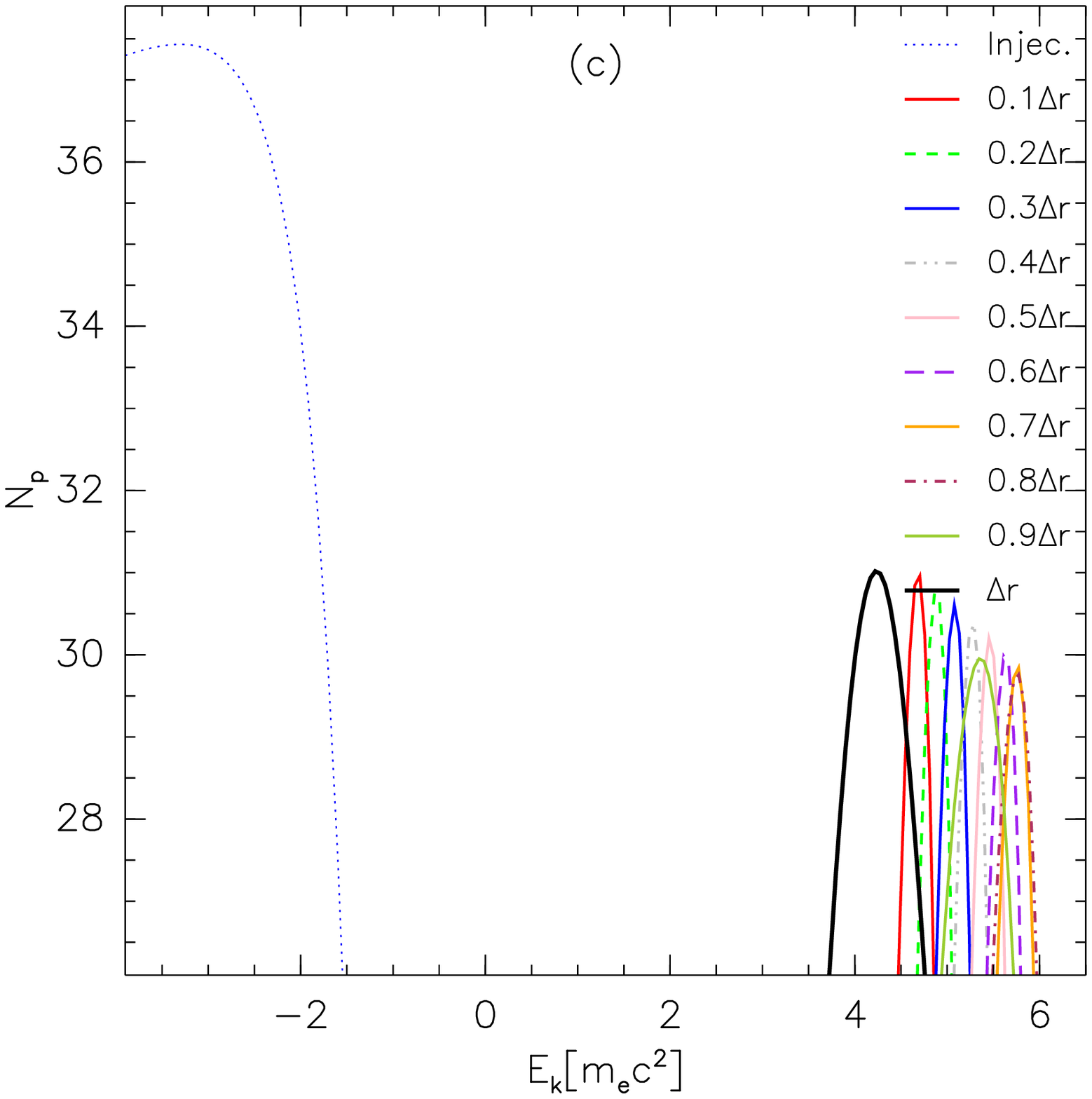}
\end{tabular}
  \end{center}
\caption{Electron spectral distributions for different MHD turbulence types: the spectral slope $\alpha=2$ for panel (a), $\alpha=5/3$ for panel (b), and $\alpha=3/2$ for panel (c). The $\delta(r_{\rm 0})$ injection source at $r_{\rm 0}$ is considered in this case.  Spectral distributions are plotted in a logarithmic interval of $0.1\Delta r$ between the curves with $\Delta r=3$. The used parameters are listed in Tables \ref{Table:fixed} and \ref{Table:free}. }  \label{figs:turindex}
\end{figure*}

Here, we explore how the spectral slope of magnetic turbulence influences the acceleration of electrons. The numerical results are plotted in Fig. \ref{figs:turindex}, where the spectral slope $\alpha=2$ is for panel (a), $\alpha=5/3$ for panel (b), and $\alpha=3/2$ for panel (c).
The $\delta(z_{\rm 0})$ function injection at $r_{\rm 0}=10^4R_{\rm g}$ is considered in this scenario. The model parameters we adopted are listed in Tables \ref{Table:fixed} and \ref{Table:free}.

As shown in Fig. \ref{figs:turindex}, spectral distribution plotted in panel (a) is the same as the panel (a2) of Fig. \ref{figs:deltayesno}, and has been explained above. Panel (b) shows that the electron acceleration with a hard turbulence index of 5/3 would produce a Gaussian-like distribution. It should be noticed that when comparing with panel (a), electrons obtain a more efficient acceleration for a small turbulence slope. As plotted in panel (c), electrons can be accelerated to a much higher energy (e.g. $E_{\rm k}\sim10^6$) for a harder turbulence index of 3/2, compared with panels (a) and (b). In a word, a hard slope of magnetic turbulence can help to efficiently accelerate low-energy electrons. Thus, we infer that the very high-energy $\gamma$-ray emission region should be located in a highly turbulent environment of the jet, and the slope of magnetic turbulence should approach the Kraichnan spectrum, 3/2.

\subsection{Influence of cooling on acceleration efficiency}\label{ICAE}

\begin{figure*}
  \begin{center}
  \begin{tabular}{cccc}
\hspace{-0.79cm}
     \includegraphics[width=55mm,height=45mm]{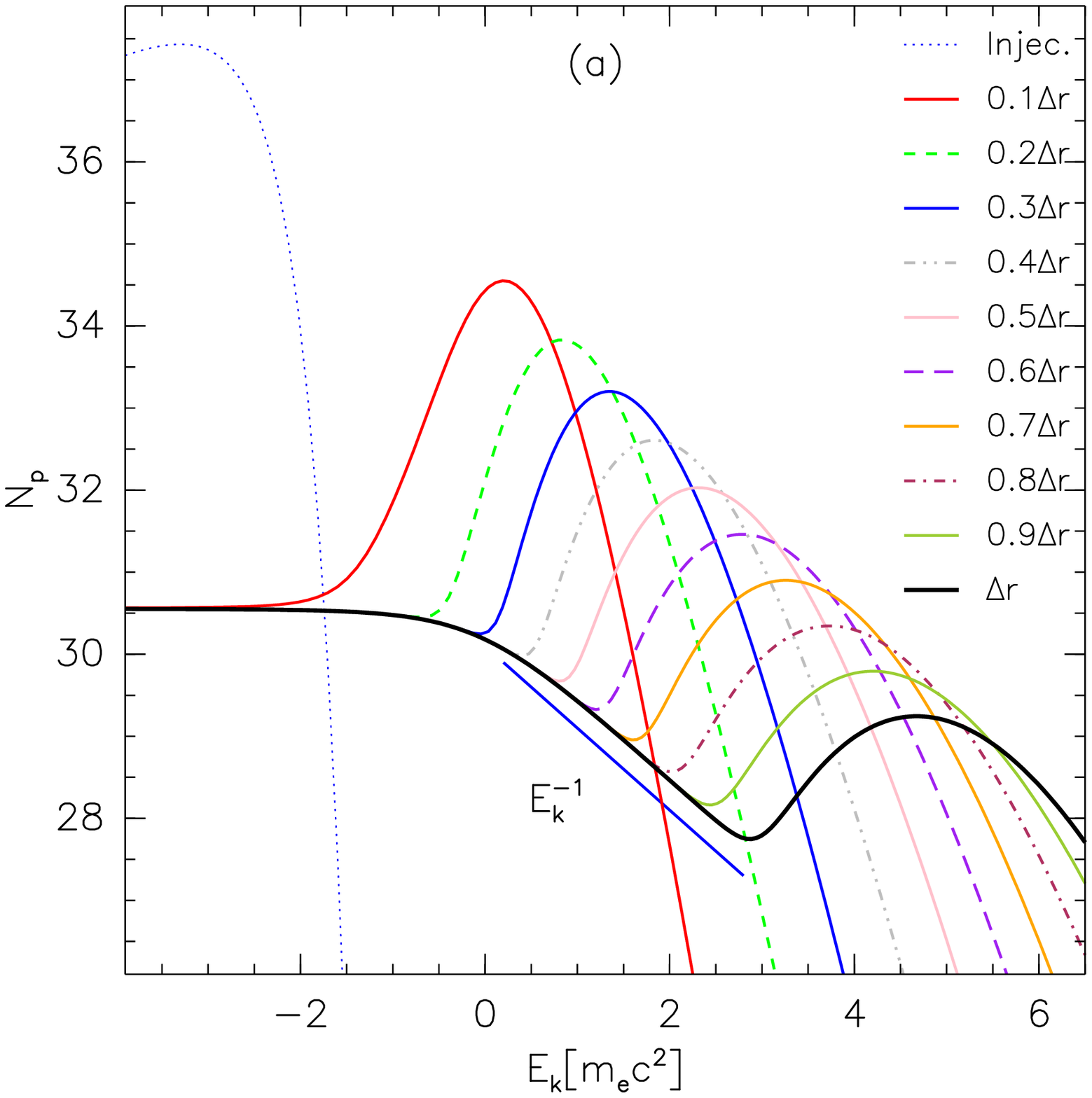}& \ \ \ \ \
\hspace{-0.79cm}
     \includegraphics[width=55mm,height=45mm]{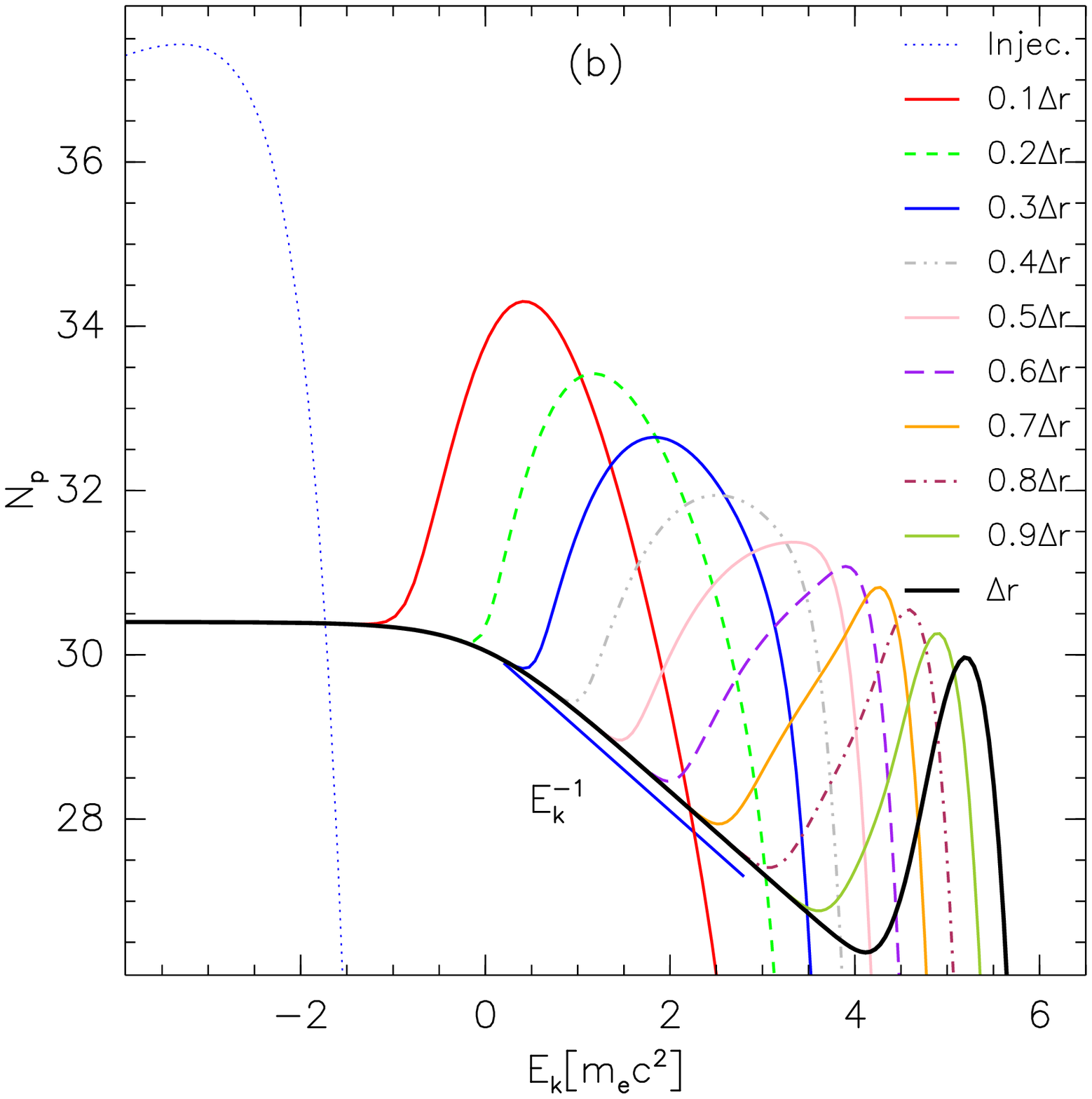}& \ \ \ \ \
\hspace{-0.79cm}
     \includegraphics[width=55mm,height=45mm]{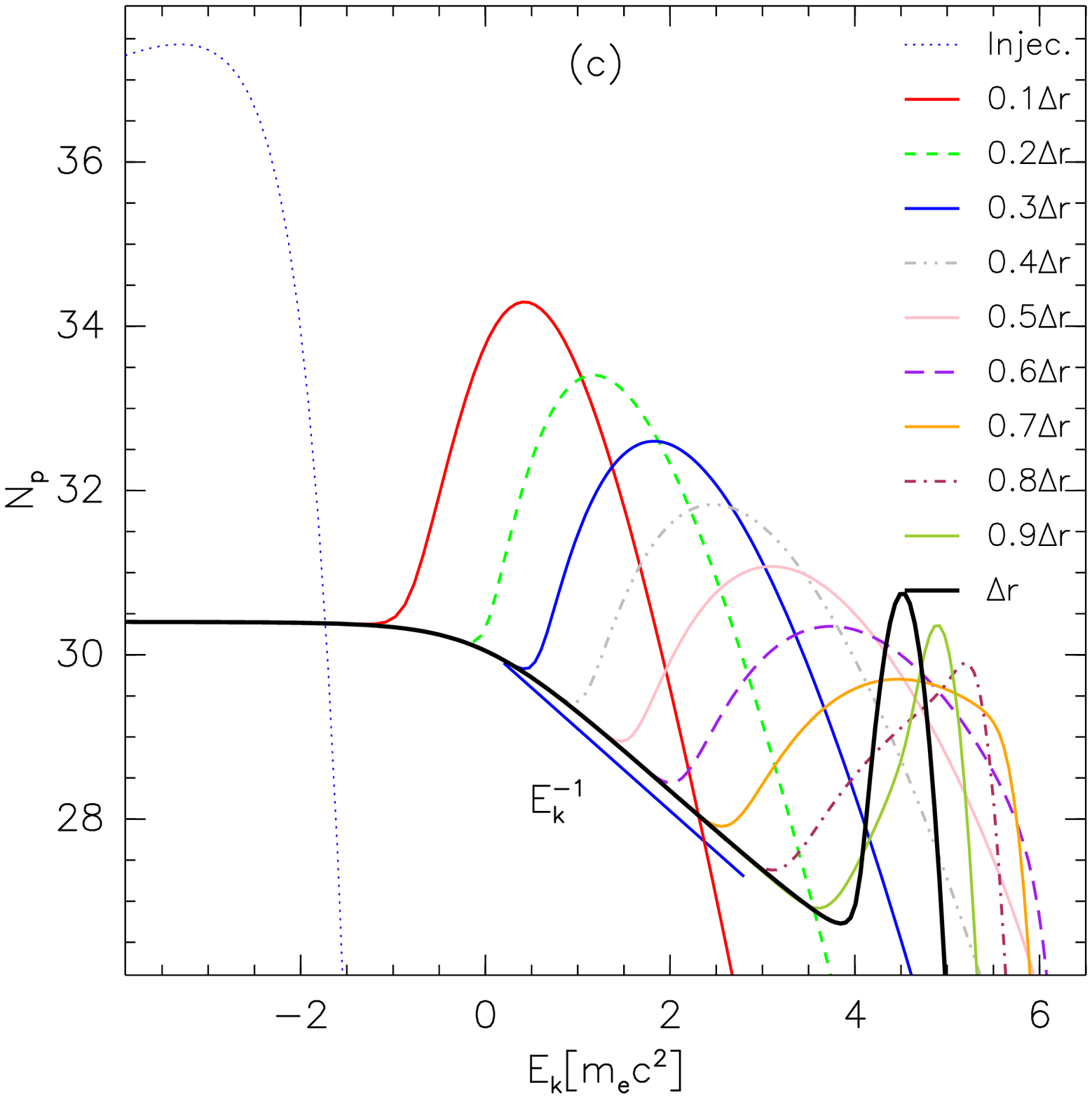}\\
     \hspace{-0.79cm}
     \includegraphics[width=55mm,height=45mm]{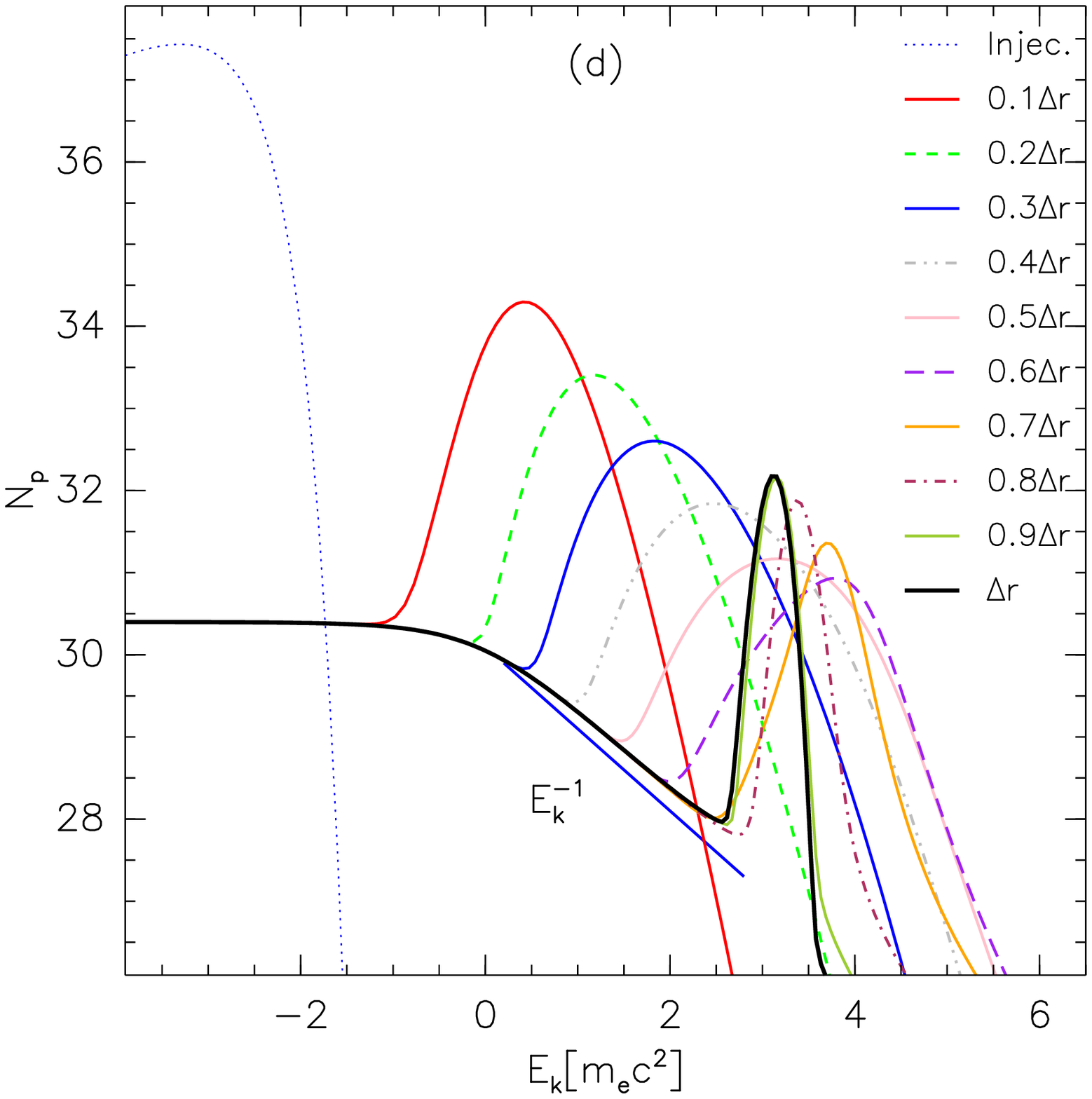}& \ \ \ \ \
\hspace{-0.79cm}
     \includegraphics[width=55mm,height=45mm]{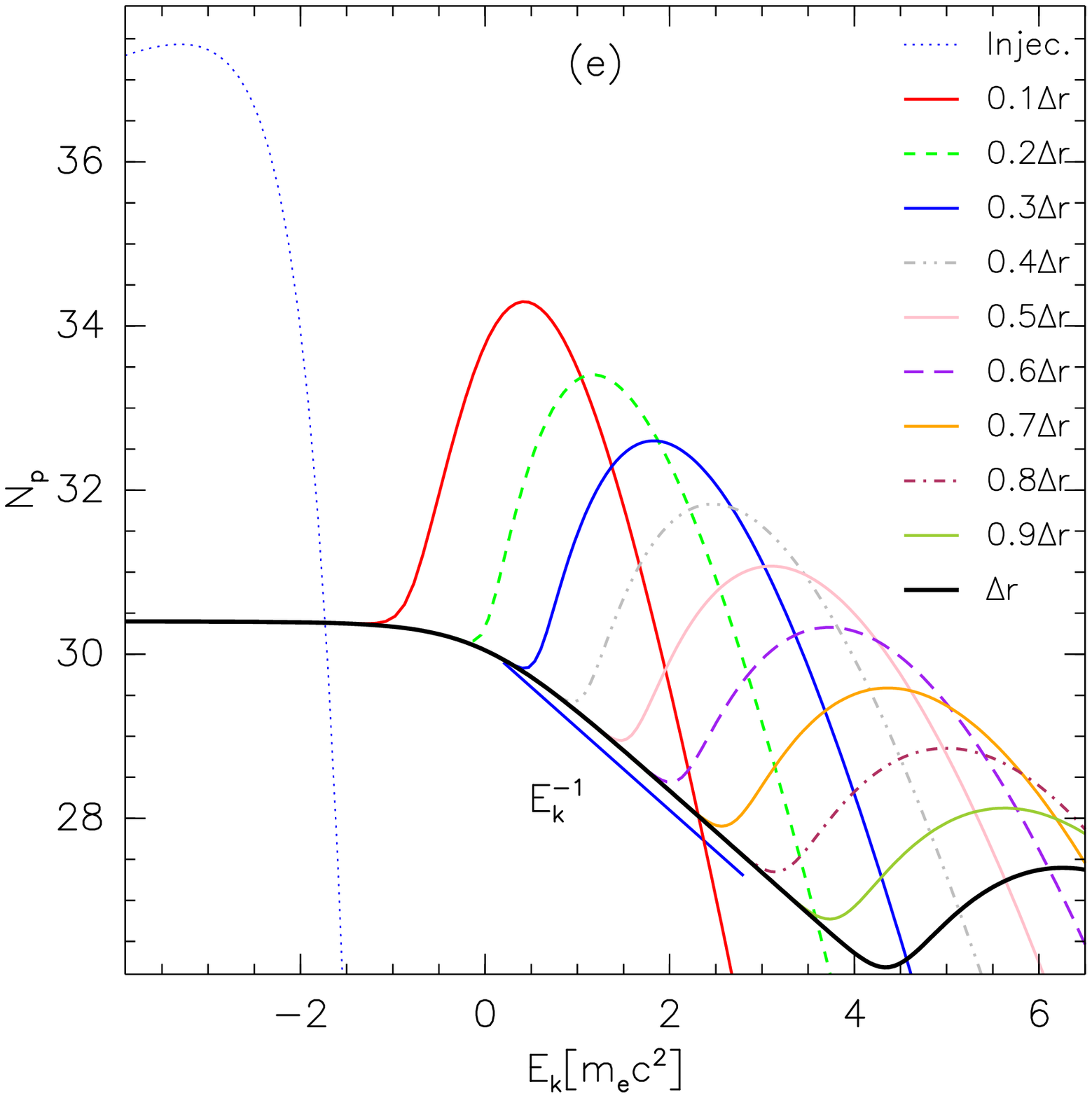}& \ \ \ \ \
\hspace{-0.79cm}
     \includegraphics[width=55mm,height=45mm]{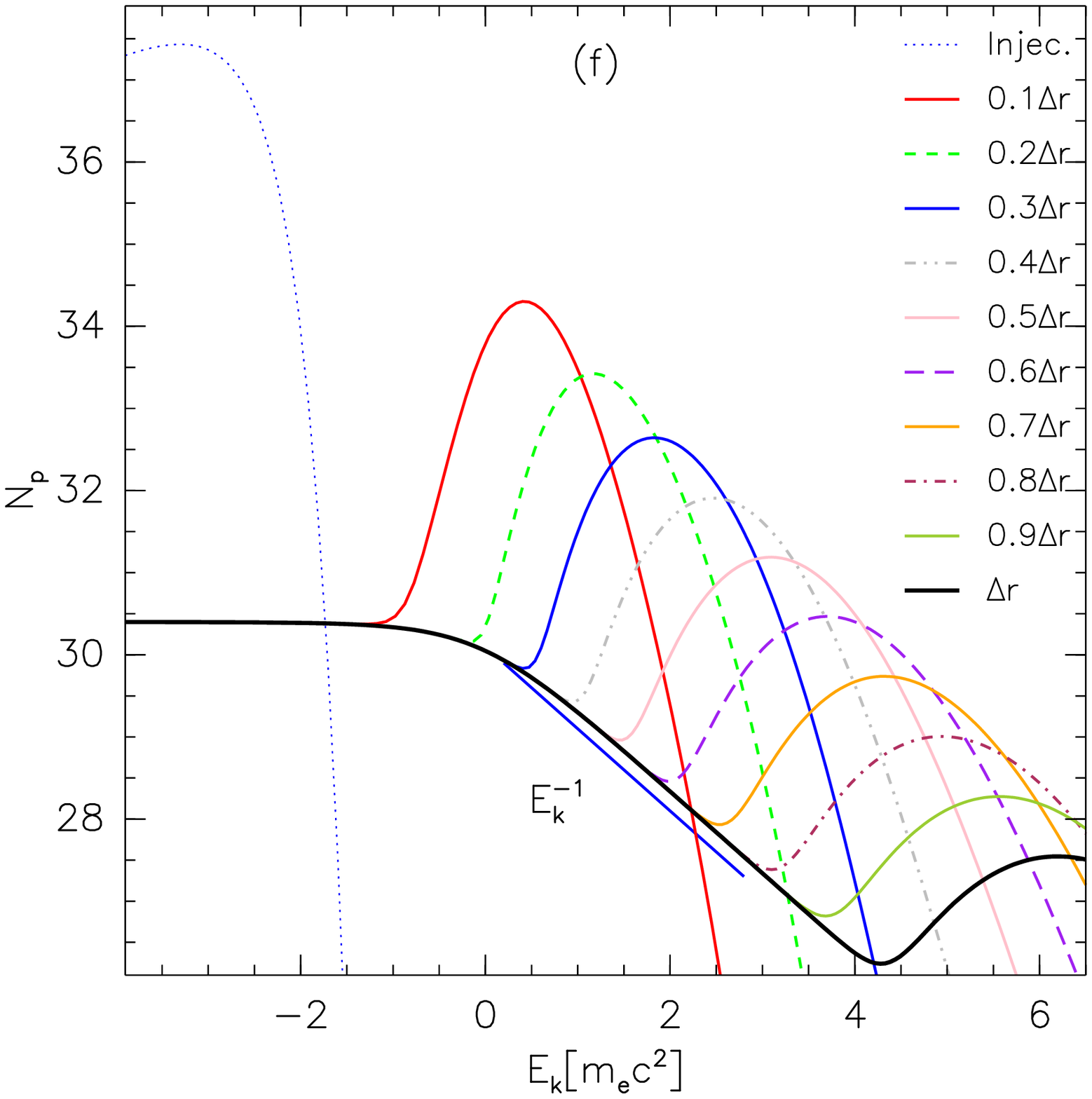}
     \end{tabular}
  \end{center}
\caption{Electron spectral distributions for various cooling rates. The influence of individual cooling rate is plotted in each panel: adiabatic cooling for panel (a); synchrotron radiation for panel (b); synchrotron self-Compton scattering for (c); ICS cooling rates of companion star photons for (d), of accretion disc photons for (e), and of background thermal photons for (f).  The $\delta(r_{\rm 0})$ function injection source is taken into account in this scenario. Electron spectral distributions are plotted in each panel in a logarithmic interval of $0.1\Delta r$ between the curves with $\Delta r=3$. The adopted parameters are listed in Tables \ref{Table:fixed} and \ref{Table:free}.}  \label{figs:radiation}
\end{figure*}

In order to study influence of various cooling rates on electron accelerations, we in Fig. \ref{figs:radiation} plot the numerical results for individual cooling mechanism, including adiabatic (panel a), synchrotron (panel b), its self-Compton scattering (panel c), ICS of companion star photons (panel d), ICS of accretion disc photons (panel e), and ICS of background thermal photons (panel f).

 The inner and outer boundaries of the acceleration region are located at $r_{\rm 0}=10^4R_{\rm g}$ and $r_{\rm end}=10^7R_{\rm g}$, respectively. The thermal electrons with the Maxwell distribution are injected only at $r_{\rm 0}$, and the other parameters are listed in Tables \ref{Table:fixed} and \ref{Table:free}.  Fig. \ref{figs:radiation}a shows the influence of adiabatic loss on spectral distributions of relativistic electrons, in which the acceleration results in a series of Planck-like spectral distributions at high-energy regime plus a power-law distribution (with index 1) with the broadening energy range at low-energy regime. It can be seen that the adiabatic loss has an impact on the spectra at $E_{\rm k}>10^3$ energy ranges, and is not enough to prevent an acceleration of thermal electron population. From Fig. \ref{figs:radiation}b, it can be found that the synchrotron loss can significantly impede the further acceleration of relativistic electrons with kinetic energy $>5\times10^5$. As shown in panels (c) and (d), both synchrotron self-Compton scattering and ICS of companion star are efficient cooling mechanisms, and can suppress the acceleration of electrons with kinetic energy $>10^6$. Moreover, they result in an evident change of spectral distributions at high-energy range due to strong radiative cooling, at the outer part of the acceleration zone. The influences of ICS of both disc and background thermal photons on the resulting spectra are studied in panel (e) and (f), respectively. From these two panels, we know that they cannot restrain an acceleration of electrons and change spectral shape.

As a result, the contributions from synchrotron, its self-Compton scattering and ICS of the stellar photons are dominant cooling channel of the accelerated electrons. These dominant cooling mechanisms can prevent not only the acceleration advance but also change spectral energy distributions of the accelerated electrons. Therefore, the acceleration efficiency of an electron depends on its cooling rate in the surrounding physical environment.

\subsection{Photon spectral energy distributions}\label{ESD}
\begin{figure*}
  \begin{center}
  \begin{tabular}{cccc}
\hspace{-0.79cm}
     \includegraphics[width=55mm,height=50mm]{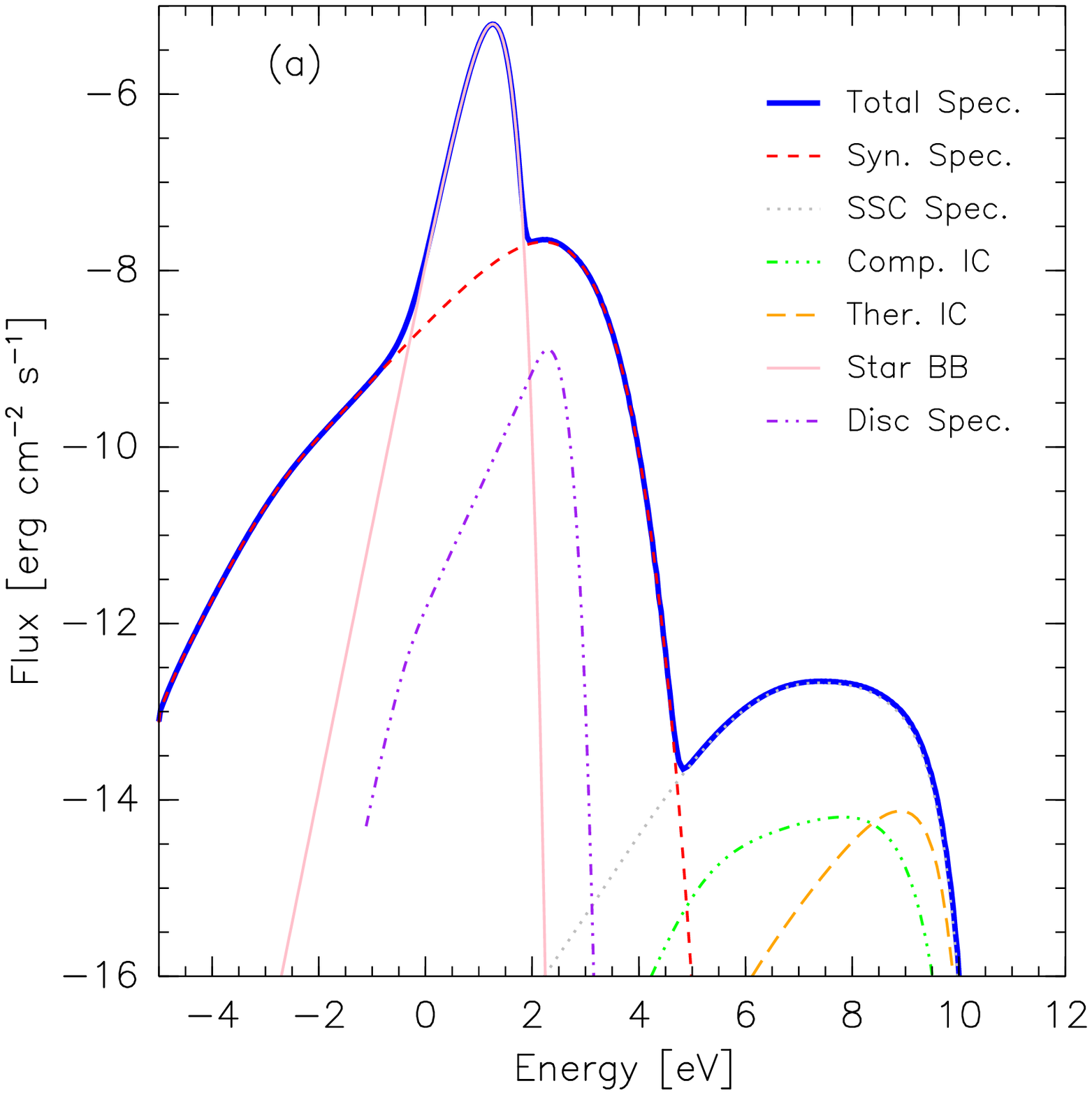}& \ \ \ \ \
\hspace{-0.79cm}
     \includegraphics[width=55mm,height=50mm]{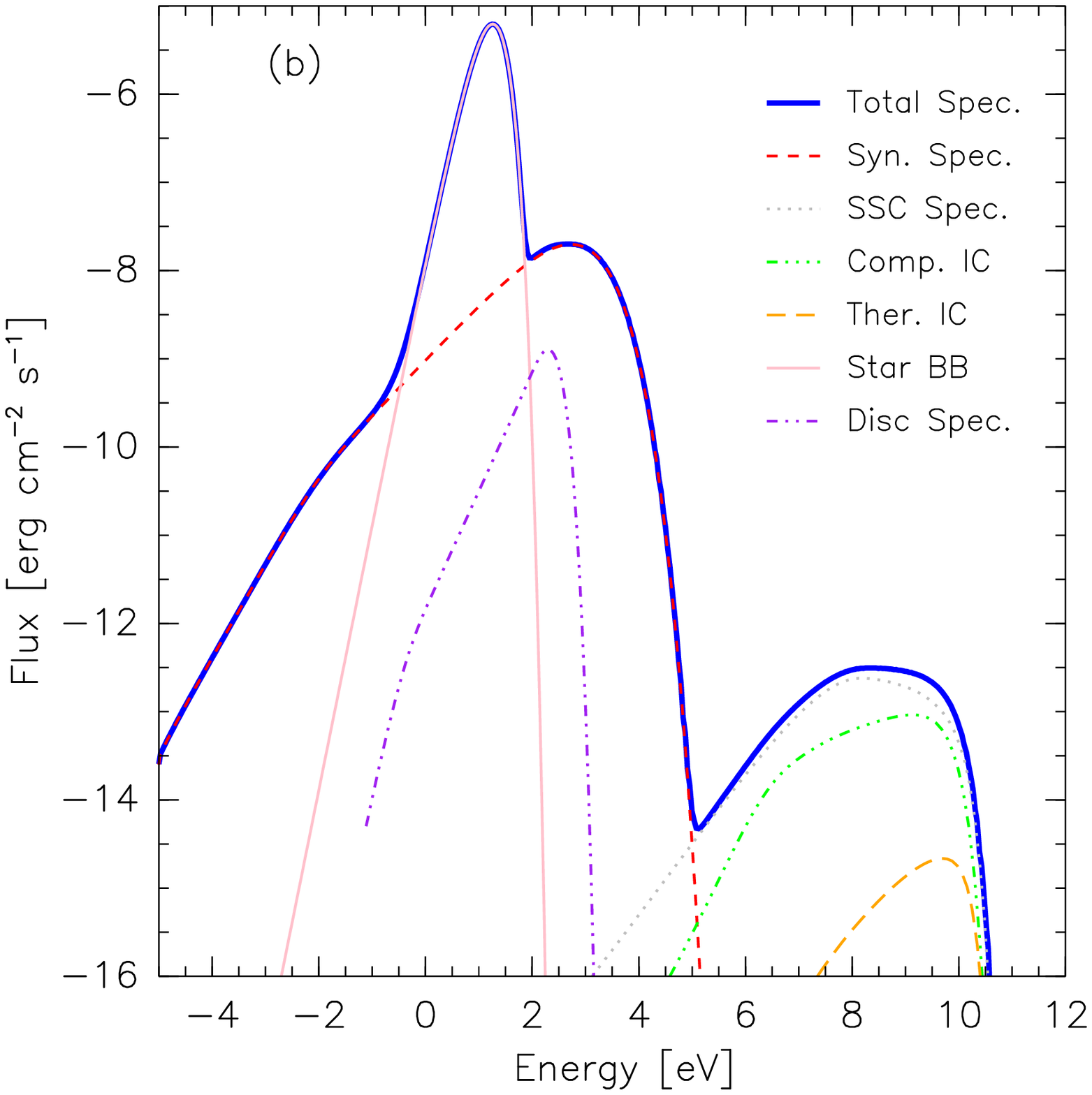}& \ \ \ \ \
\hspace{-0.79cm}
     \includegraphics[width=55mm,height=50mm]{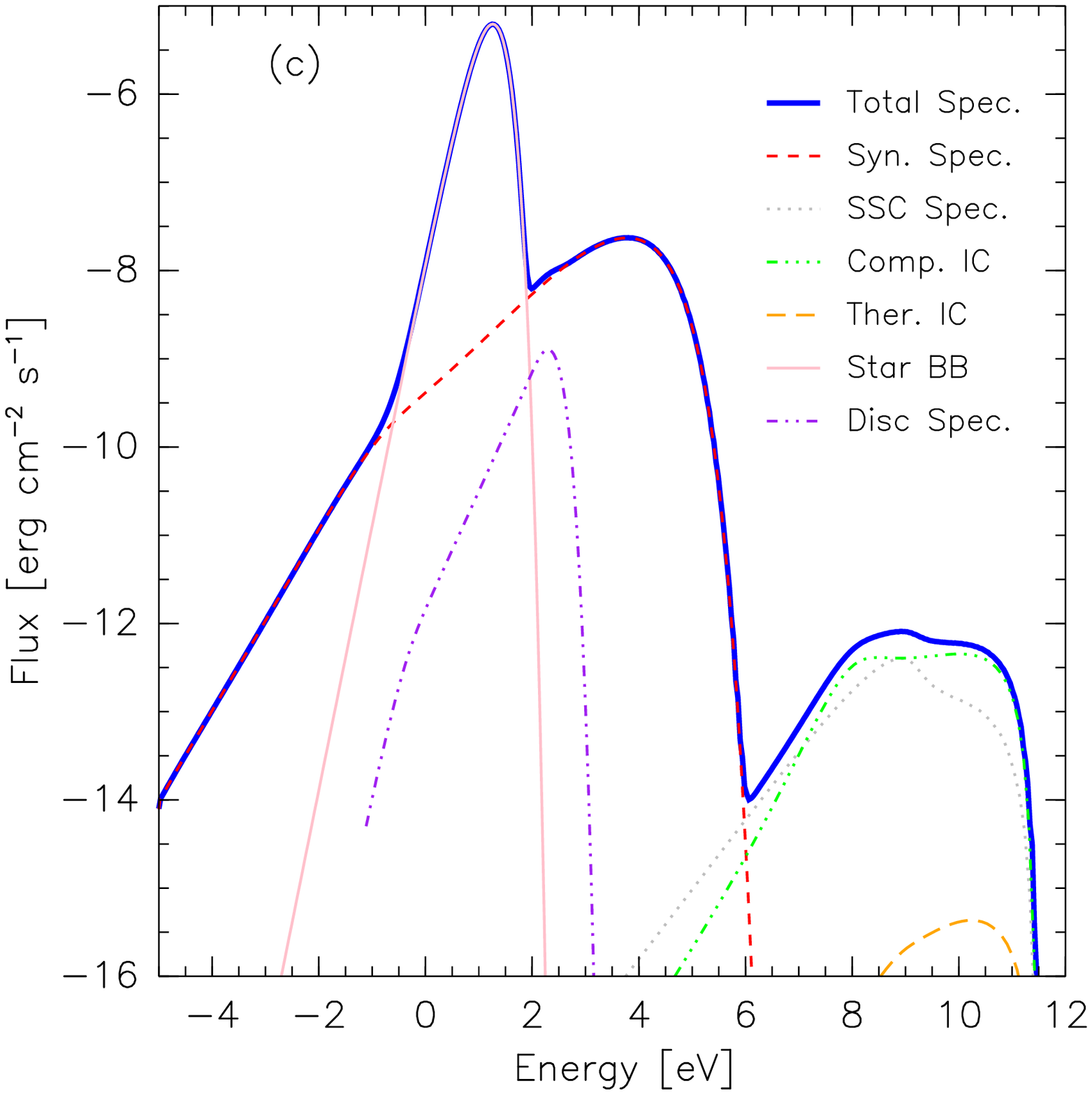}
     \end{tabular}
  \end{center}
\caption{Multi-wavelength spectral energy distributions with different ratios of the magnetic energy density to the jet matter one: $\xi=10$ for panel (a), $\xi=1$ for panel (b) and $\xi=0.1$ for panel (c).  Legends indicating \emph{Total Spec.}: total output spectrum; \emph{Syn. Spec.}: synchrotron radiation spectrum; \emph{SSC Spec.}: synchrotron self-Compton scattering spectrum; \emph{Comp. IC}: ICS spectrum from the companion photons; \emph{Ther. IC}: ICS spectrum from the background thermal photons; \emph{Star BB}: blackbody spectrum of the companion; \emph{Disc Spec.}: multi-temperature blackbody spectrum of the accretion disc. Both ICS of accretion disc photons and relativistic bremsstrahlung are negligible and not presented in this figure.The total SEDs plotted does not consider absorption interactions of high-energy photons by soft photon fields. The adopted parameters are listed in Tables \ref{Table:fixed} and \ref{Table:free}. } \label{figs:spectra}
\end{figure*}

After obtaining spectral distributions of accelerated electrons, we further study the interactions of these electrons with the ordered magnetic fields and surrounding soft photon fields. As a matter of fact, synchrotron emission from electrons interacting the ordered magnetic fields has been computed in advance in order to explore the influence of synchrotron self-Compton scattering cooling rate on acceleration processes. By using parameters listed in Tables \ref{Table:fixed} and \ref{Table:free}, we study the multi-wavelength spectral energy distributions via considering the case of the continuous injection of electrons with a thermal Maxwell distribution along the extended acceleration region. Numerical results are demonstrated in Fig. \ref{figs:spectra}: the case of magnetic pressure dominance for panel (a); an equipartition case between magnetic pressure and jet matter one for panel (b);  and the scenario of the jet matter density dominance for panel (c).

As shown in Fig. \ref{figs:spectra}, synchrotron radiation dominates the emission output at radio, IR and soft X-ray wavebands. Soft photon fields from the companion star provide fluxes at visible light range. Furthermore, the synchrotron self-Compton scattering dominates emission fluxes ranging from hard X-ray to high-energy $\gamma$-ray bands. For the case of considering the strong magnetic energy density in the jet (panel a), where a shock is mediated by strong magnetic pressure, ICS spectra of both the companion star and background thermal photons are negligible. When decreasing magnetic energy density, i.e. for the equipartition (panel b) and sub-equipartition (panel c) cases, ICS contributions from the companion star photons increase. For the latter, the shock interaction dominates the acceleration of electrons. For panel (c), high-energy radiation is from synchrotron self-Compton scattering and ICS of background thermal photons. For the currently adopted parameters, in particular, the location of the acceleration and emission regions being away from the central compact object, relativistic bremsstrahlung radiation and ICS from accretion disc photons are negligible and not shown in this figure.

We emphasize that the spectra presented in Fig. \ref{figs:spectra} are subjected to the used parameters, but it is a common fact that total non-thermal emission spectra are composed of a synchrotron spectrum at the low energy regime and synchrotron self-Compton and/or external Compton scattering at the high-energy regime.

\section{Discussions}
MHD turbulence is considered as an important agent in both stochastic and shock interaction processes (e.g. \citealt{Petrosian12}). For the former, i.e. second-order \emph{Fermi} process, interactions of random fluctuation magnetic turbulence with particles lead to an increase of their energies and their spectral distribution changes. Regarding the latter, magnetic turbulence is also a key ingredient for accelerated particles repeatedly crossing the shock surface. Our studies demonstrated that for the turbulent magnetized X-ray binary jet, dominant magnetic turbulence with a hard spectral slope can more effectively accelerate electrons. Besides, we have also explored the fact that competing interaction between acceleration rates and cooling losses can change spectral distributions and suppress an increase of energies of relativistic particles. However, this work did not explore how the properties of magnetic turbulence affect the output photon spectral distributions. It will be an interesting topic for studying the radiative properties of accelerated particles in a turbulent environment mixing some large- and small-scale magnetic fields

Recently, by assuming an anisotropic configuration of turbulent magnetic fields embedded in the X-ray binary jets, \cite{Zhang17} studied the properties of polarized radiation produced in jets. On the basis of this work, these turbulent magnetic fields were regarded as a mixing component of large- and small-scale configurations. When the characteristic length of turbulence is less than the non-relativistic Larmor radius, magnetic turbulence results in a polarized jitter radiation, which has a higher peak frequency than the synchrotron one (\citealt{Kelner13}). On the contrary, the polarized synchrotron emission occurs, which is similar to the usual synchrotron radiation produced by the ordered, large-scale magnetic fields (\citealt{Kelner13,Prosekin16}). It needs to further consider the jitter radiation spectra and polarization properties in the current model. Thanks for having obtained new electron distributions from a viewpoint of the first principle in this work, a further study can lead to a self-consistent in determining both the polarization radiation and the origin of the particles.

Although this paper focus only on acceleration processes of electrons, we emphasize that numerical techniques provided in this work could also apply heavy element accelerations. In the framework of hadronic model, studying proton and electron simultaneous accelerations will be helpful for understanding the properties of MHD turbulence, matter compositions of the jet, as well as radiation and polarization characteristics.  After obtaining broadband spectral energy distributions of a general microquasar (see Fig. \ref{figs:spectra}), an inertia of thoughts would usually apply the model to fit observations from one or a few object sources, in order to test the feasibility of the model. However, this paper mainly focused on the acceleration aspects of electrons and has studied the influence of MHD turbulence on electrons' energization, from the whole viewpoint of microquasar class.

In Sect. \ref{ESD}, we have given multi-broadband spectral energy distributions for different ratios between magnetic energy density and jet matter one, from a general microquasar. Our work did not take into account an electromagnetic cascade process of pair creation due to absorption interaction of high-energy photons by several photon fields. On the basis of the condition of pair creation, i.e. $\geq m_{\rm e}^2c^4/\epsilon$, one knows that the absorption of TeV photons is influenced by the surrounding companion star photon fields with the energy $\epsilon$, and the absorption process at GeV ranges is due to photons of accretion disc,  and background thermal photons. In general, when a dissipation region is close to the central compact object, absorption interactions from star and disc photon fields would become more significant (\citealt{Bednarek93,Romero10b,Cerutti11}). If one considers a cascade process, the spectra at the GeV range plotted in Fig. \ref{figs:spectra} would be attenuated moderately (e.g. \citealt{Zhang14}); meanwhile, pair creation would induce a secondary radiative process (e.g. \citealt{Zhang10}), which produces an increase of radiation flux at low-energy regime.

In general, an absorption of photons is dependent on the viewing angle and the orbital phase, so the SEDs shown in Fig. \ref{figs:spectra} are just indicative of the total photon production. This result would be modified by absorption due to both intrinsic to the jet and in the external fields,
and cascades if ICS dominates the radiative losses. However, cascades can be cut by the synchrotron channel when it dominates under the condition of the strong magnetic field strength.

As for the class of high-mass microquasars, the central black hole is accreting the companion matter mainly via stellar wind to power the system. The massive stellar companion provides a seed photon field (with a few $\rm eV$ energy) of ICS, radiation photons from which can help us to understand the radiative properties of the system. Due to the cooling of relativistic electrons by this channel, acceleration of electrons would be suppressed (see Sect. \ref{ICAE}). According to the studies in Sect. \ref{ICAE}, other radiation channels such as synchrotron, and its self-Compton scattering that are also very important cooling channels can heavily impede an acceleration of electrons. In order to explain high-energy observations of microquasars by \emph{Fermi} LAT or AGILE, the maximum energy the accelerated electrons require is $\gamma\sim 10^5$, by synchrotron self-Compton or external Compton scattering channels. The results we present in Sect. \ref{NumRes} demonstrate that for some common parameter choices such as turbulence index $5/3$, a dominant turbulent magnetic field $\zeta$, and an accelerator location $r_{\rm 0}>10^4R_{\rm g}$ in the jet, both stochastic and shock interactions can accelerate low-energy (background thermal) electrons up to the required value $\gamma\sim10^5$. Furthermore, we find that if a microquasar system can emit very high-energy $\gamma$-ray signal, which need provide the ultra-relativistic electrons with energy $\gamma>10^6$, magnetic fields of the jet are dominated by turbulent fluctuation component with the spectral slope close to the fast mode, corresponding to the spectrum of acoustic turbulence with the index $\alpha=3/2$, i.e. the well-known Kraichnan spectrum.

Regarding the class of low-mass microquasars, the seed photon field from the companion star is negligible during the low/hard spectral states, in which synchrotron self-Compton scattering and ICS of thermal photons are dominant radiation channels at the high-energy or very high-energy bands (see \citealt{Zhang15}). The maximum energy of photons observed is $400\rm \ MeV$ in V404 Cygni (\citealt{Loh16,Piano17}), which corresponds to the need for $\gamma\sim10^4$ electrons via the dominant radiative process. On the basis of Sect. \ref{NumRes}, we find that for some common parameters choices the accelerated mechanism related to magnetic turbulence can easily accelerate electrons up to $\gamma\sim10^4$. However, due to dominant cooling processes (such as synchrotron, its self-Compton scattering, ICS of thermal photons) competing with acceleration rates from \emph{Fermi} I and II mechanisms, it needs to provide some extreme parameters such as a large $\zeta$ and a small turbulence index $\alpha$ to make acceleration efficient enough. Therefore, we infer that the class of low-mass microquasars is some effective GeV $\gamma$-ray radiative sources.

Besides studies on the maximum energy the electron acceleration can reach to, spectral energy distribution of accelerated particles is also attracting much attention, which is related to the classical injection problems in the particle acceleration theories.  In Sect. \ref{EIS}, we have explored three scenarios of an electron injection source: Maxwell, Gaussian, and power-law distributions. We find that the former two types usually give the same spectral distributions of the accelerated electrons. In particular, the latter (power-law injection) also produces a similar distribution but presents a power-law tail (for the case of continuous injection, see Fig. \ref{figs:deltayesno}c1). Thus, it can be seen that the resulting spectra do not always keep a common power-law form of non-thermal electrons.  In the other words, non-thermal electrons accelerated in jets present a Maxwell-like or Gaussian-like (i.e. thermal-like) distribution.

\section{SUMMARY}
In this paper, we have studied acceleration processes of relativistic electrons by stochastic and shock mechanisms in the turbulent magnetized jet environment of X-ray binaries. The numerical results we have obtained are briefly summarized as follows.

1. MHD turbulence plays a critical role in the acceleration processes of relativistic particles in the jets via stochastic diffusion and shock collisions.
In the case of magnetic pressure dominated, the stochastic acceleration process is more efficient than the shock collision interactions, and vice versa.

2. The dominant turbulent magnetic field is very necessary for an efficient acceleration of relativistic electrons. Furthermore, the spectral slope of MHD turbulence can significantly influence the relativistic electron acceleration, in which the hard spectral index, such as Kraichnan spectrum 3/2, corresponds to a more efficient acceleration.

3. The spectral energy distributions of the relativistic electrons do not depend on the given initial injection spectra of electrons. In general, the new non-thermal electron population produced during acceleration processes demonstrates a thermal-like distribution form.

4. Ultra-relativistic particle's accelerator should be located at the distance $>10^3R_{\rm g}$ away from the central black hole. Both \emph{Fermi} I and II acceleration rates competing with the various cooling rates result in changes of electron spectral distribution and an increase of their energies.

5. The acceleration mechanisms studied in this paper can naturally accelerate electrons up to the energy $\gamma\sim10^4$ required for explaining GeV observations of microquasars. However, in order to predict very high-energy $\gamma$-ray observations of microquasars, it needs to  adopt some relatively extreme parameters to enhance accelerator abilities.

\section*{ACKNOWLEDGMENTS}
We would like to thank an anonymous referee for his/her valuable comments and suggestions that helped to improve this manuscript significantly.
We thank also financial support provided by the National Natural Science Foundation of China under grants 11703020, 11233006, 11363003, U1531108, and U1731106. J.F.Z. acknowledges support from the research project of Xiangtan University under grant No. 10KZ/KZ08089.
%\input{Zhang.tex}

%\label{lastpage}

\end{document}